\documentclass[traditabstract]{aa}
\usepackage{txfonts}
\usepackage{natbib}
\usepackage{graphicx}
\usepackage{dblfloatfix}

\bibpunct{(}{)}{;}{a}{}{,} 

\newcommand{\kms}{km~s$^{-1}$\,}

\newcommand{\ndeg}{^{\circ}}

\newcommand {\sL}{\rm\,L$_\odot$}
\newcommand {\mgi} {Mg\,{\small I}\,}
\newcommand {\caii} {Ca\,{\small II}\,}
\newcommand {\vi} {{\it V--I\/}}
\newcommand {\mv} {{\it V\/}}

\begin{document}

\title{Cleaning spectroscopic samples of stars in nearby dwarf galaxies\thanks{Based on FLAMES observations collected at the ESO, proposals 171.B-0588, 076.B-0391, 079.B-0435}}
\subtitle{The use of the nIR Mg\,{\small I}\, line to weed out Milky Way contaminants} 

\author{Giuseppina Battaglia\inst{1,2} \and Else Starkenburg\inst{3}}

\institute{
European Organization for Astronomical Research in the Southern Hemisphere, K. Schwarzschild-Str. 2, 85748 Garching, 
Germany\label{inst1} 
\and
INAF - Osservatorio Astronomico di Bologna, via Ranzani 1, 40127, Bologna, Italy \email{gbattaglia@oabo.inaf.it}\label{inst2} 
\and 
Kapteyn Astronomical Institute, University of Groningen, PO Box 800, 9700 AV Groningen, The Netherlands \email{else@astro.rug.nl}\label{inst3} }

\date{Received / Accepted}

\abstract{
Dwarf galaxies provide insights on the processes of star formation and chemical enrichment at the low end of the galaxy mass function, as well as on the clustering of dark matter on small scales. In studies of Local Group dwarf galaxies, spectroscopic samples of individual stars are used to derive the internal kinematics and abundance properties of these galaxies. It is therefore important to clean these samples from Milky Way stars, not related to the dwarf galaxy, since they can contaminate the analysis of the properties of these objects. Here we introduce a new diagnostic for separating Milky Way contaminant stars -- that mainly constitute of dwarf stars -- and red giant branch stars targeted in dwarf galaxies. As discriminator we use the trends in the equivalent width of the nIR \mgi line at 8806.8 \AA\, as a function of the equivalent width of \caii triplet lines. This method is particularly useful for works dealing with multi-object intermediate resolution spectroscopy focusing in the region of the nIR \caii triplet.  We use synthetic spectra to explore how the equivalent width of these lines changes for stars with different properties (gravity, effective temperature, metallicity) and find that a discrimination among giants above the horizontal branch and dwarfs can be made with this method at [Fe/H]$> -2$ dex. For $-2 \le$ [Fe/H] $\le -1$, this method is also valid to discriminate dwarfs and giants down to approximately one magnitude below the horizontal branch. Using a foreground model we make predictions on the use of this new discrimination method for nearby dwarf spheroidal galaxies, including the ultra-faints. We subsequently use VLT/FLAMES data for the Sextans, Sculptor and Fornax dSphs to verify the predicted theoretical trends.
}


\titlerunning{The nIR Mg\,{\small I}\, line to weed out Milky Way contaminants}
\authorrunning{G. Battaglia \& E. Starkenburg}

\maketitle

\section{Introduction}

There is much to be learned from the galaxies within the Local Group which can be studied in better detail than any other system, on a star-by-star basis. Within the Milky Way (MW) halo itself, we find satellite galaxies which span a range over 10 magnitudes in brightness. The great majority of the MW satellites belong to the class of early type dwarf galaxies, which are the smallest and least luminous galaxies known to date. These include the ultra faint dwarfs (UFDs) which were relatively recently discovered with the Sloan Digital Sky Survey (SDSS). Several future surveys, like Pan-STARRS \citep{kais02} and Skymapper \citep{kell07}, are expected to further increase the number of detected dwarf galaxies around the MW, towards lower surface brightnesses.

Great effort is devoted to the study of these small early type galaxies \citep[see][for a review]{tht09} as they provide precious insights on the processes that drive galaxy evolution at the low end of the galaxy mass function, as well as on the clustering of dark matter on small scales \citep[e.g.][]{gilm07}. Since stars in these galaxies are resolved, the internal properties for the galaxies are derived from photometric and spectroscopic studies of individual objects. This requires the application of methods to discard from the sample foreground/background stars which are not bound to the dwarf galaxy, but that are instead within the MW itself (hereafter ``contaminants'' or ``interlopers''), and that would influence the properties derived for the dwarf galaxy. 

General methods to exclude obvious interlopers include the use of color-magnitude diagrams (CMDs) to select a region in magnitude and color where one would expect member stars in a certain stellar evolutionary stage to lie given the distance to the object. For most of the studies of Local Group early type dwarf galaxies a photometric selection box is placed on the red giant branch (RGB) - the most luminous feature in the CMD of ancient/intermediate age stellar populations. However, such a selection box will also contain contaminants, these consist predominantly of much intrinsically fainter but close-by main sequence dwarf stars from the MW thin or thick disk. In order to distinguish the contaminant dwarf stars from the RGB stars members to the dwarf galaxy additional discriminators need to be used.

Spectroscopic studies can rely on a further selection criterion to weed out foreground contamination, by using the line-of-sight velocity of the individual stars as compared to the systemic velocity of the dwarf galaxy. This is commonly done in a statistical sense, either using sigma-clipping procedures (which imply a negligible probability to find a member beyond a certain velocity threshold) or maximum likelihood analyses, with more detailed implementations modeling both the expected population of dwarf galaxy stars and MW stars \citep[e.g.][]{batt08scl, walk09a, mart11}. Additional properties than line-of-sight velocities can be included in the analysis, e.g. distance of the star from the galaxy center \citep[e.g.][]{batt08scl}, and information on spectral indices \citep[][for the Mg index around $\lambda$ 5170 \AA\,]{walk09a} or metallicity \citep{mart11} of the star. The most sophisticated of these implementations from a statistical point of view are those of \citet{walk09a} and \citet{mart11}.

However, even the most efficient of these methods cannot by construction select out MW contaminants with similar properties (colors, apparent magnitudes, line-of-sight velocities etc) as the member stars to the observed galaxy. Depending on the type of stars surveyed, the systemic velocity of the galaxy and its position on the sky (most importantly Galactic latitude), contaminant stars satisfying all these requirements might still be numerous or at least statistically significant in number. Furthermore, a selection criterion based on position, velocity (and/or metallicity) needs assumptions about the underlying distribution and can therefore also bias the final analysis. For example, it is unclear whether features in the outer parts could be mistaken for contaminants: if looking for tidally disrupted stars, one expects to find them in the very outer parts of the dwarf and at velocities rather discrepant from the systemic \citep[e.g.][for an observational study of tidal disruption in Carina]{muno06}. In statistical procedures those stars would most likely be considered as non-members. Similarly, stars in the outer ``shell'' of the Fornax dSph \citep{cole05}, arguably the remnant of an accretion event, would most likely to rejected because well outside the limiting radius of the dwarf as inferred by its surface brightness profile. Also, stars with very low and high metallicity with respect to the rest of the population, could be rejected as non-members, while true members with deviating metallicities could give important insights on the earliest phases of chemical enrichment and on the capability of the dwarf to retain gas and/or metals, respectively. 

These examples illustrate the benefits of having also methods that weed out contaminants on the basis of sensitivity to gravity, since the stars targeted in external galaxies are RGB stars and the foreground consists mostly of dwarf stars from the thin or thick disk. Such a discriminator can be used separately or combined with other (statistical) methods to increase the efficiency of the elimination of interlopers. 

Photometric techniques are present in the literature which allow a distinction between dwarf and giant stars by constructing indices sensitive to gravity and luminosity from combinations of particular filters. For example, the DDO51 filter is centered on the Mg b/MgH feature near 5170 \AA\, , which is strong in late-type dwarfs and weak in giants (e.g. Morrison et al. 2001). Several works in the literature have shown the effectiveness of separating giant and dwarf stars using a combination of Washington and DDO filters: \citet[see e.g.][]{muno06} pre-selected photometrically their spectroscopic targets using the M51 filter, finding probable members to the Carina dSph out to very large distances from its center, where genuine members would have otherwise been buried in the foreground. \citet{fari07} used Str\"{o}mgren photometry on the Draco dSph, showing  that the Str\"{o}mgren $c_1$ index, sensitive to luminosity, can be used together to the $(b-y)_0$ color to separate stars in various evolutionary stages. While photometric techniques provide broad information on the star formation history of the dwarf galaxy and in some cases on the metallicity distribution of its stars \citep[e.g.][]{fari07}, information on the line-of-sight velocity of the individual stars appears crucial to assess membership in some cases even when using Washington photometry \citep[e.g.][]{morr01}.

There also exist spectroscopic studies of early type Local Group dwarf galaxies that have used features in the spectra of the stars to discriminate between giant and dwarfs. \citet{spin71} and \citet{schi97} showed the Na~I doublet of absorption lines at 8183, 8195 \AA\, to be strongly dependent on gravity and temperature. Several studies have adopted the equivalent width (EW) of the Na~I doublet as a diagnostic to discriminate between dwarf and giant stars, with the dwarf stars displaying much larger EWs than giant stars \citep[see][on the halo of M31 for the former, and on UFDs for the latter two works]{gilb06, simo07, mart07}. In their work, \citet{gilb06} showed that this diagnostic offers clear differentiation at colors (V-I)$ > $ 2, while there appears to be significant overlap in the distribution of Na~I doublet EW of dwarf and giant stars for bluer colors. 

The adoption of one, or more, method to assess membership depends on the particular scientific problem, galaxy, and data-set under consideration, and the fraction of contaminants that can be tolerated so as not to significantly affect the results. In some cases the presence of interlopers has a dramatic impact on the conclusions. For example mass estimates derived from measurements of the line-of-sight velocity dispersion of samples of individual stars in LG early type dwarf galaxies have been used to discuss the possible existence of ``universal'' mass value \citep[e.g.][]{stri08, walk09b}. In their study of UFDs, \citet{simo07} adopted a wealth of membership indicators, such as spatial position of the star, location on the CMD, EW of the Na~I doublet, line-of-sight velocity and metallicity, and found for the Hercules dSph an internal velocity dispersion of 5.1$\pm$0.9 km/s. By removing MW stars using Str\"omgren photometry, \citet{aden09a} found a lower velocity dispersion in their cleaned sample, 3.7$\pm$0.9 \kms. In this particular example, the derived mass of \citet{aden09b} from this new velocity dispersion places Hercules well below the ``common mass scale'' for MW early type dwarf galaxies \citep[][and references therein]{stri08}.

Another instructive example is the case of Willman~1, whose apparent spread in the metal abundance distribution was used in favor of the classification as a very low luminosity dwarf galaxy \citep{mart07}. Subsequent studies based on high resolution spectroscopic observations revealed that a large percentage of stars classified as members on the basis of their position on the CMD, velocity and EW of the Na~I doublet by \citet{mart07} were either foreground dwarfs or halo giants, renewing the possibility that Willman~1 could be a disrupted metal-poor globular cluster.

The examples above show the importance of careful interloper removal by using as many membership indicators as possible. This is particularly important as the galaxies studied become fainter and consist of less stars, and for studies which explore the outer regions of dwarf galaxies, where the expected ratio of members/interlopers is large. 

\begin{figure}[]
\begin{center}
\includegraphics[width=0.9\linewidth]{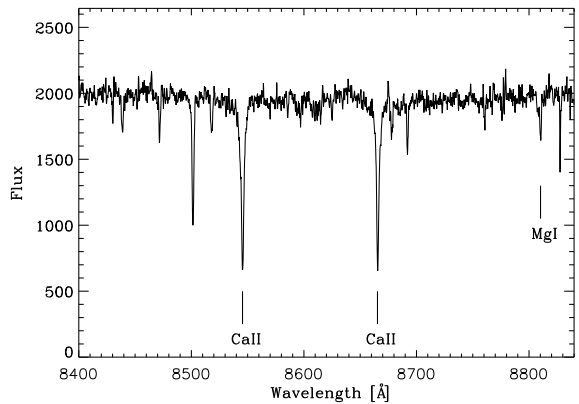}
\caption{VLT/FLAMES spectrum from DART observations of a star along the line-of-sight to the Sculptor dSph 
(heliocentric velocity $=$ 122.7 \kms and signal-to-noise ratio S/N/\AA\, $=$64). The lines used in this work, i.e. 
the two strongest CaT lines and one \mgi line, are marked.  \label{fig:spectrum_scl}}
\end{center}
\end{figure}

In this paper we present an additional discriminator to separate dwarf and giant stars, i.e. the EW of the Mg~I line at 8806.8 \AA\, used in combination with the sum of the EW of the two strongest \caii triplet (CaT) lines ($\Sigma$W), at $\lambda=$ 8542.1 and 8662.1 \AA\, (see Figure~\ref{fig:spectrum_scl}). This method was empirically introduced by \citet[][hereafter B11]{batt11} as a criterion to discriminate RGB stars belonging to the Sextans dSph from the MW contaminant stars. Using a data-set with a limited extent in metallicity, the authors noted that the great majority of the stars with line-of-sight velocities very close to the systemic of Sextans (i.e. highly likely member RGB stars) clustered around small values of the Mg~I EW, while stars with velocities more than 4$\sigma$ away from the systemic (i.e. highly likely non-members) clustered at larger values, $>$ 0.5 \AA\,. In this paper we extend the empirical work of B11: we use synthetic spectra to explore the validity of the method over a range of metallicities, gravities and $\Sigma$W, demonstrating the effectiveness of using the Mg~I line in combination to the CaT lines in removing MW contaminants. 

This method is a convenient choice: \mgi and CaT lines are close in wavelength space and relatively broad, which means they will typically both be present and measurable in intermediate-resolution studies of this wavelength regime. They are located in the red part of the spectrum, where the target RGB stars are brightest, with optimization of the exposure time. Furthermore, the wavelength region including CaT lines has become increasingly more used in intermediate resolution spectroscopic studies of LG early type dwarf galaxies: CaT lines are both very suitable for radial velocity determinations and offer the possibility of deriving estimates of [Fe/H] accurate to $\pm$0.15-0.2 dex over a wide metallicity range \citep{star10} for large numbers of individual stars and much lower observing times with respect to high resolution studies (typically a S/N$ \sim 10$ is required). The method can be applied to a wealth of galaxies observed in this wavelength regime. In the ESO archive a large body of data already exists at intermediate resolution in the only set-up including the CaT, which covers the region between 8206 and 9400 \AA. This is by far the most frequently used set-up for the FLAMES instrument. Note that the Na~I doublet is just outside this wavelength range. The current archive includes observations of Fornax, Sculptor, Sextans, Carina, Leo~II, Leo~IV, Hercules, Boo~I, IC~1613, Sag, Segue~1 in this wavelength range, by various groups. The CaT region has also been targeted by studies using other telescopes than VLT, such as the work of Koch et al. 2007 on Leo~I. Furthermore, the future large GAIA-ESO Survey (GES) will target the wavelength region covering the nIR CaT and Mg~I line (V. Hill, GES Team, private communication). Several existing and future data-set could therefore benefit from this discriminator.

The paper is structured as follows. In Sect.~\ref{sec:lines} we use synthetic spectra to explore the trends between the EW of \mgi line and CaT lines for stars of different metallicity and gravity and to derive a relation that allows us to discriminate among giant and dwarf stars on the basis of these lines. We then comment on the expected success rate of this method along the line-of-sight to different MW satellites in Sect.~\ref{sec:foreg}. In Sect.~\ref{sec:obs} we use data for the Sextans, Sculptor and Fornax dSphs from the Dwarf Abundances and Radial Velocities Team \citep[DART,][]{tols06}, to explore the behavior of the lines empirically for galaxies with different characteristics, compare the results to the models and show the agreement between models and observations. Finally we apply the method to the DART data-set in Sect.~\ref{sec:clean} and present our conclusions in Sect.~\ref{sec:conc}.

\section{The lines and their sensitivity} \label{sec:lines}

\begin{figure}
\includegraphics[width=\linewidth]{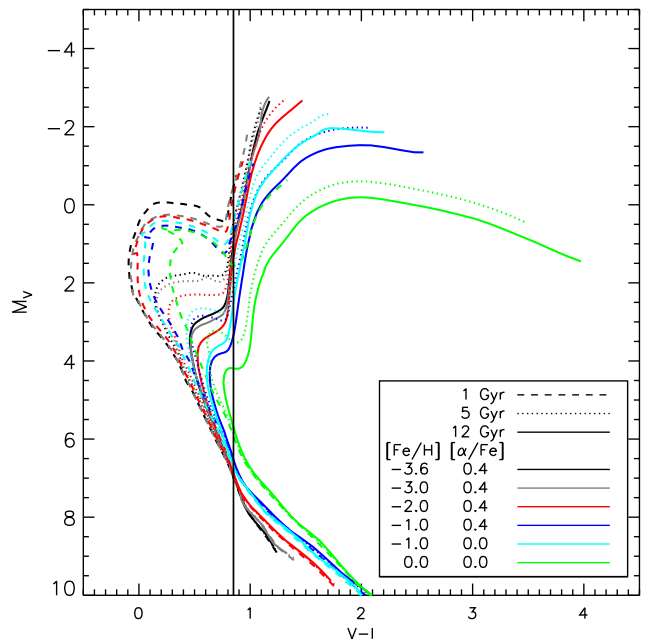}
\caption{Colour-magnitude diagram using Yonsei-Yale isochrones for 1 Gyr (dashed lines), 5 Gyr (dotted lines), and 12 Gyr (full lines) and metallicities [Fe/H]=-3.6 (lowest value possible within the Yonsei-Yale set in black), [Fe/H]=-3.0 (grey), [Fe/H]=-2.0 (red), [Fe/H]=-1.0 (blue), [Fe/H]=0.0 (green) and [Fe/H]=+0.5 (purple). [$\alpha$/Fe] is set to +0.4 for all isochrones except for [Fe/H]=0.0 and [Fe/H]=+0.5 which have [$\alpha$/Fe]=0.0 and an extra set of isochrones plotted for [Fe/H]=-1 and [$\alpha$/Fe]=0.0 (cyan). The vertical line roughly indicates the onset of the RGB. \label{CMDs}}
\end{figure}

In order to investigate the behavior of the nIR CaT and \mgi absorption lines we first have to define the boundaries of the physical parameter space we are working in. In practise, we do this using isochrones covering possible ranges of [Fe/H] and ages as covered by the stars in Local Group dwarf galaxies and MW foreground in order to select the appropriate regions in log(g) - T$_{\textnormal{\scriptsize{eff}}}$ space (see Figure~\ref{CMDs}). The isochrones we use are Yonsei-Yale isochrones \citep[e.g.][]{yi01,dema04} for 1, 5  and 12 Gyr, covering the wide range $-3.6 \le$ [Fe/H] $\le +0.0$. To represent general trends within the MW stars, the alpha enhancements chosen are +0.4 for [Fe/H]$\le -1$ dex and +0.0 for the solar [Fe/H]. However, since spectroscopic  observations of stars in the Sculptor \citep[e.g.][]{tht09}, Fornax \citep{leta10} and Carina \citep{koch08} dSphs show [$\alpha$/Fe] around solar values already at [Fe/H]$\sim -1$ dex, for completeness we add an isochrone set of [$\alpha$/Fe]=0.0 dex and [Fe/H]$=-1$ dex.

 Because the dwarf galaxies currently studied in large spectroscopic surveys \citep[e.g][]{walk06,simo07,aden09a,walk09a, kirb10} are distant objects, typically only the giants above the horizontal branch (HB) will be bright enough to be followed up with spectroscopy; in systems with sparsely populated RGBs though observations often need to be pushed to lower magnitudes, therefore below the HB \citep[e.g.][]{mart07, simo07}. The isochrones plotted in Fig.~\ref{CMDs} show that giant stars of any metallicity or age above the HB \citep[which will have $0<$M$_{V}<+1$ typically][]{rich05,cate08}, and most of those below the HB, will not be found 
at bluer colours than V-I $\sim$0.85, corresponding to stars hotter than $\sim$5250 K \citep{rami05}. In our following analysis, aiming at disentangling the populations of giant and dwarf stars which would be pre-selected photometrically, we therefore will not take temperatures $>$ 5250 K into account. This temperature selection cuts out an important fraction of the hot, halo and thick disk stars in the foreground contamination. Figure \ref{CMDs} also shows that at the lowest metallicity end the isochrones become less sensitive to metallicity and thus we expect no significant changes if even lower metallicities ([Fe/H]=-4 dex and below) would be considered in the analysis.

On the other end of the temperature scale we place a cut-off at V-I$=2.1$, corresponding to a temperature of 3600K. Although in theory very metal-rich giants stars can reach even cooler temperatures, these are rare in dwarf galaxies because most dwarf galaxies have low metallicity. Indeed, following this consideration, most photometric boxes that have been used for pre-selection of spectroscopic targets are well within this range. Note that with this cut, the only population of giant stars completely excluded from the analysis is the one of super-solar metallicity and with an age of 12 Gyr, unexpected in such systems. 

Since we address in this work the validity of the use of the \mgi and CaT lines to discriminate RGB stars in Local Group dwarf galaxies from MW contaminants, we use in the analysis synthetic spectra with physical properties (log(g), T$_{\textnormal{\scriptsize{eff}}}$, etc) covering the range expected for both populations of stars within the color range discussed above. We have calculated a grid of synthetic spectra for a wide range of metallicities, temperatures and gravity using (OS)MARCS model atmospheres \citep[e.g.][]{gust08,plez08} and the Turbospectrum program \citep{alva98}, updated consistently with the \citet{gust08} MARCS release. In agreement with the trends followed by most of the stars in the MW \citep[e.g.][]{venn04}, we adopt MARCS models with ``standard'' chemical composition, i.e. the values chosen for the alpha elements (taken as O, Ne, Mg, Si, S, Ar, Ca, and Ti) relative to Fe for the models are [$\alpha$/Fe]=+0.4 at the metallicities below solar, linearly decreasing towards [$\alpha$/Fe]=0.0 at solar [Fe/H]. The microturbulent velocities used in our models are 1--2 \kms, in agreement with the observed range \citep[e.g.][]{bark05}.

\begin{figure*}[]
\includegraphics[width=\linewidth]{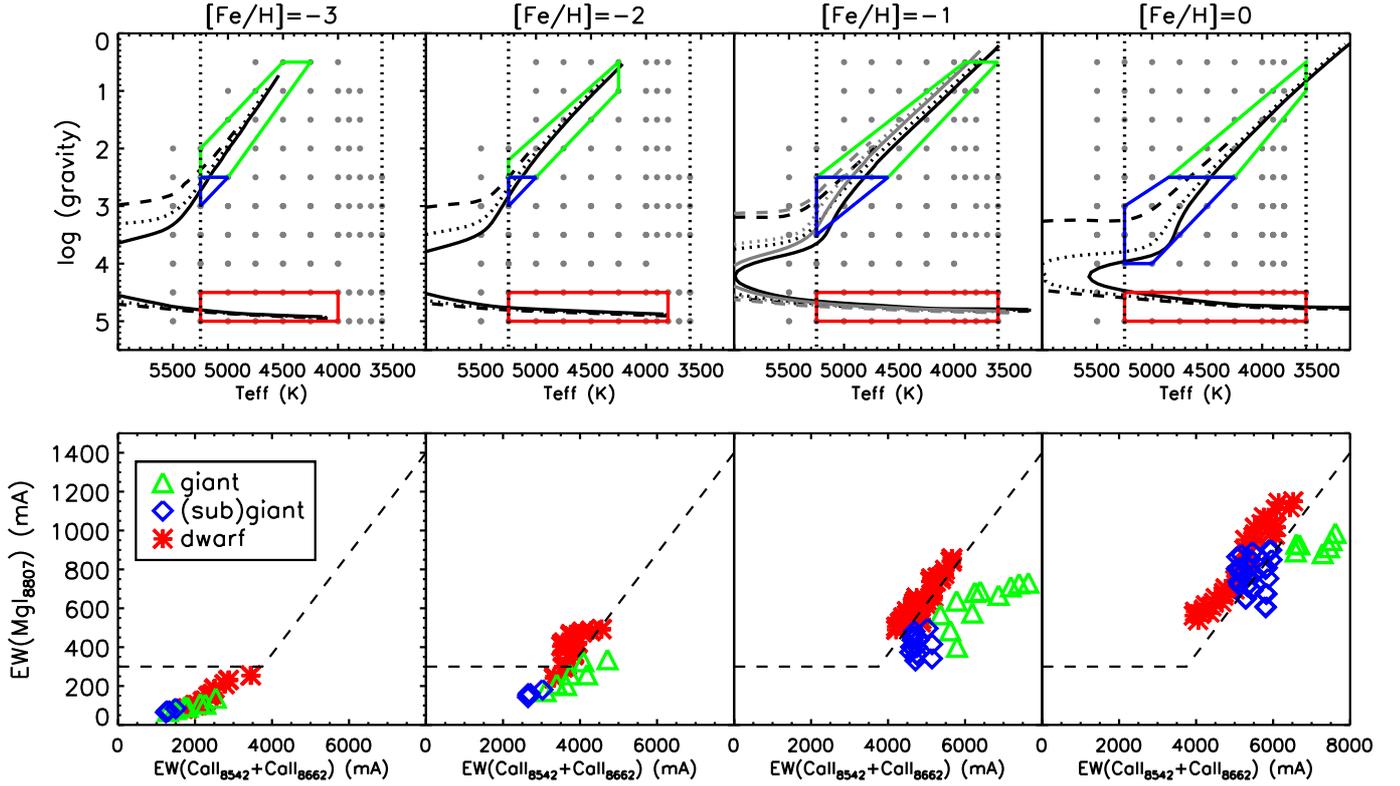}
\caption{Upper panels: Yonsei-Yale isochrones for 1 Gyr (dashed lines), 5 Gyr (dotted lines), and 12 Gyr (full lines) for one particular metallicity. For [Fe/H]=-3 and [Fe/H]=-2, [$\alpha$/Fe]=0.4, for [Fe/H]=0.0, [$\alpha$/Fe]=0.0 and for [Fe/H] both alpha enhancements are given ([$\alpha$/Fe]=0.0 in grey). All values for which the synthetic spectra are calculated are covered with a grey dot. Overplotted are regions where the giants above the HB (green), (sub)giants below the HB (blue) and main-sequence (red) branches would lie in Teff-log(g) space, if the temperature restrictions from photometry are taken into account (vertical dotted lines). Lower panels: EWs for the two strongest CaT line versus the MgI 8806.8 line for models that fall within the boxes of the panel above. Giants above the HB are shown as green triangles, (sub)giants below the HB as blue diamonds and dwarfs as red asterisks. Overplotted a distinction line for dwarfs and giants (except for the extremely low-metallicity case), this line is the same in all panels.\label{iso}}
\end{figure*}

To allow a comparison of the data with the synthetic spectra we define regions in T$_{\textnormal{\scriptsize{eff}}}$-log(g) space where the giants above the HB, (sub)giants below the HB and main-sequence branches would lie (see upper panels of Figure \ref{iso}). For ``classical'' dSphs, which have a well populated RGB, spectroscopic samples mostly target RGB stars above the HB, therefore the member stars we are looking for will all fall within the 'giant' boxes (green boxes in Figure \ref{iso}). For those UFDs for which observations must be pushed below the HB, members will also fall within the 'sub-giant' boxes. For all the models that fall within either one of the selection boxes we measure the EW of the \mgi line and of the two strongest CaT lines in the corresponding synthetic spectrum. We measure the EW of the CaT lines using a Gaussian fit to the individual lines, corrected by a factor 1.1 to include the non-Gaussian wings. This factor is determined by comparing the EW from the Gaussian fit to the one from the integrated line measure (for more details see Battaglia et al. \citeyear{batt08} and Starkenburg et al. \citeyear{star10}). We also correct the EWs of the CaT lines for non-LTE effects using the equations given in \citet{star10}, these corrections are mainly important for the very low-metallicity regime. The \mgi line is measured using a simple integration of the area under the continuum of 6 \AA\, around the line, as also done for the data (see Sect.~\ref{sec:obs}).

In the bottom panel of Figure \ref{iso}, we show the sum of the EW for the two strongest CaT lines ($\Sigma {\rm W}_{\rm CaT}$) versus the EW of the \mgi line (${\rm EW}_{\rm Mg}$). For [Fe/H]$=-2$ and $-1$ dex the dwarf stars can be separated from the giants and sub-giants. At solar metallicity, the distinction remains clear between the dwarfs and the giants, but not for the sub-giants. Note though that, given the metallicity-luminosity relation \citep[see][for early type dwarf galaxies in the Local Group]{kirb10}, stars with metallicities close to solar are not expected in UFDs, which are the systems for which one often needs to target stars below the HB due to their sparsely populated RGB. The figure also shows that the distinction among the different classes of objects would not be possible if one were to collapse all data onto one axis (i.e. would use just one of the lines): it is the combination of these lines that makes the distinction between the giants and dwarfs possible in most cases. In the extremely low-metallicity regime (i.e., [Fe/H]$=-3$), giants, sub-giants and dwarfs appear to occupy the same region in the ${\rm EW}_{\rm Mg}$, $\Sigma {\rm W}_{\rm CaT}$ plane. However this is not a concern because in Sect.~\ref{sec:foreg} - where we discuss the expected contamination for different objects - we will show that interloper stars with such low metallicities are very rare.

A distinction line between giants and dwarfs is overplotted in all panels and is:
\begin{equation} \label{eq:line} 
{\rm EW}_{\rm Mg} (m\AA\,) = \left\{ \begin{array}{ll}
 300 &\mbox{ if $\lambda \le$ 3750 m\AA\,} \\
 0.26 \times \Sigma {\rm W}_{\rm CaT} -670.6  &\mbox{ if $ \lambda >$ 3750 m\AA\,}
       \end{array} \right.
\end{equation}

  As we will discuss in Sect.~\ref{sec:foreg}, the fraction and type of contaminant stars, and therefore their location and importance on the ${\rm EW}_{\rm Mg}$ vs $\Sigma {\rm W}_{\rm CaT}$ plane, will change according to the specifics of the dwarf galaxy (position on the sky, systemic velocity, distance). This line therefore provides a criterion to keep in as many of the member stars of the dwarf galaxy, and throw away as many contaminants as one can.

\begin{table*}[]
\caption{Properties of the foreground contamination towards the MW satellites. The columns list: the name of the galaxy (1), its galactic coordinates (2,3); distance modulus from \citet{mateo98} for the classical satellites, \citet{belo07} for Coma B, Hercules, CVenII, Leo IV and Segue I, \citet{zuck06a} for UMaII, \citet{zuck06b} for CVenI and \citet{simo07} for UMaI and \citet{mart07} for Willman~1 (4); 
  $V$-band magnitude of the horizontal branch from \citet{mart07} for UMaI and CVenI, \citet{aden09a} for Hercules and \citet{ih95} for the classical satellites (5); 
   heliocentric systemic velocities from B11 for Sextans, \citet{batt06} for Fornax, \citet{batt08} for Sculptor, \citet{koch06} for Carina, \citet{mateo98} for the classical dwarf galaxies and \citet{simo07} for most ultra-faint dwarf galaxies, except Segue I \citep{simo10} and Willman~1 \citep{mart07} (6); corresponding velocity dispersion from the same references except we use \citet{walk09a} for the classical dwarf galaxies with systemic velocities from \citet{mateo98} (7); the number of expected MW contaminants within a solid angle of 4 deg$^2$ along the line-of-sight to the galaxy (8);
  The last 4 columns list 
  the percentage of contaminants retained in the sample when using only the velocity criterion (9), the line criterion (10), when applying both (11) and applying both relaxing the velocity selection to 4-$\sigma$ (12). }
\label{tab:sel}
\centering
\begin{tabular}{lrrllrrrrrrr}
\hline
\hline 
Galaxy & l & b & dm & V$_{\rm HB}$ & v$_{\rm sys}$ & $\sigma$ & N$_{cont}$ & \%(N$_{cont}$) & \%(N$_{cont}$) & \%(N$_{cont}$) & \%(N$_{cont}$) \\
 & ($\ndeg$) & ($\ndeg$) & & & [\kms] & [\kms] & / 4 deg$^2$ & vel. crit. (3$\sigma$) & line crit. & both (3$\sigma$) & both (4$\sigma$) \\
\hline 
Fornax & 237.1 & --65.7 & 20.7 & 21.29 &  54.1 & 11.4 & 2309  & 44\% & 17\% & 6\% & 8\%\\
Sculptor & 287.5 & --83.2 & 19.54 & 20.13 & 110.6 & 10.1 & 1872 & 4\% & 16\% & 2\% & 3\%\\ 
Sextans & 243.5 & 42.3 & 19.67 & 20.35 & 226.0 & 8.4 & 3687 & 0.9\% & 12\% & 0.5\% & 0.6\%\\ 
\hline
Canes Venatici II & 74.3 & 79.8 & 20.9 & & --128.9 & 4.6 & 2891 & 2\% & 14\% & 0.9\% & 1\%\\
Canes Venatici I & 113.6 & 82.7 & 21.75 & 22.4 & 30.9 & 7.6 & 2283 & 21\% & 19\% & 3\% & 5\%\\
Carina & 260.1 & --22.2 & 20.03 & 20.50 & 223.9 & 7.5 & 18000 & 0.7\% & 16\% & 0.5\% & 0.7\%\\
Coma Berenices & 241.9 & 83.6 & 18.2 & & 98.1 & 4.6 & 1738 & 1\% & 11\% & 0.5\% & 0.7\% \\
Draco & 86.4 & 34.7 & 19.58 & 20.07 & --293.0 & 9.1 & 6329 & 0.5\% & 15\% & 0.4\% & 0.5\%\\
Hercules & 28.7 & 36.9 & 20.7 & 21.17 & 45.0 & 5.1 & 13368 & 13\% & 22\% & 3\% & 4\%\\
Leo I & 226.0 & 49.1 & 21.99 & 22.30 & 286.0 & 9.2 & 2617 & 1\% & 16\% & 0.3\% & 0.5\% \\ 
Leo II & 220.2 & 67.2 & 21.63 & 22.30 & 76.0 & 6.6 & 2248 & 12\% & 19\% & 3\% & 4\% \\
Leo IV & 265.4 & 56.5 & 21.0 & & 132.2 & 7.6 & 3930 &  6\% & 13\% & 2\% & 2\% \\
Segue I & 220.5 & 50.4 & 16.8 & & 208.5 & 3.7 & 1637 & 0.2\% & 9\% & 0.1\% & 0.2\% \\ 
Ursa Major I & 152.5 & 37.4 & 20.13 & 20.5 & --55.3 & 7.6 & 4096 & 24\% & 11\% & 3\% & 4\% \\
Ursa Major II & 159.4 & 54.4 & 17.5 & & --116.5 & 6.7 & 1868 & 2\% & 10\% & 0.9\% & 1\% \\
Ursa Minor & 105.0 & 44.8 & 19.11 & 19.80 & --248 & 9.5 & 3449 & 0.8\% & 14\% & 0.6\% & 0.8\%\\
Willman 1 & 158.6 & 56.8 & 17.9 & & --12.3 & 4.3 & 2026 & 32\% & 8\% & 1\% & 2\%\\ 
\hline
\end{tabular}
\end{table*}

\begin{figure*}[]
\includegraphics[width=\linewidth]{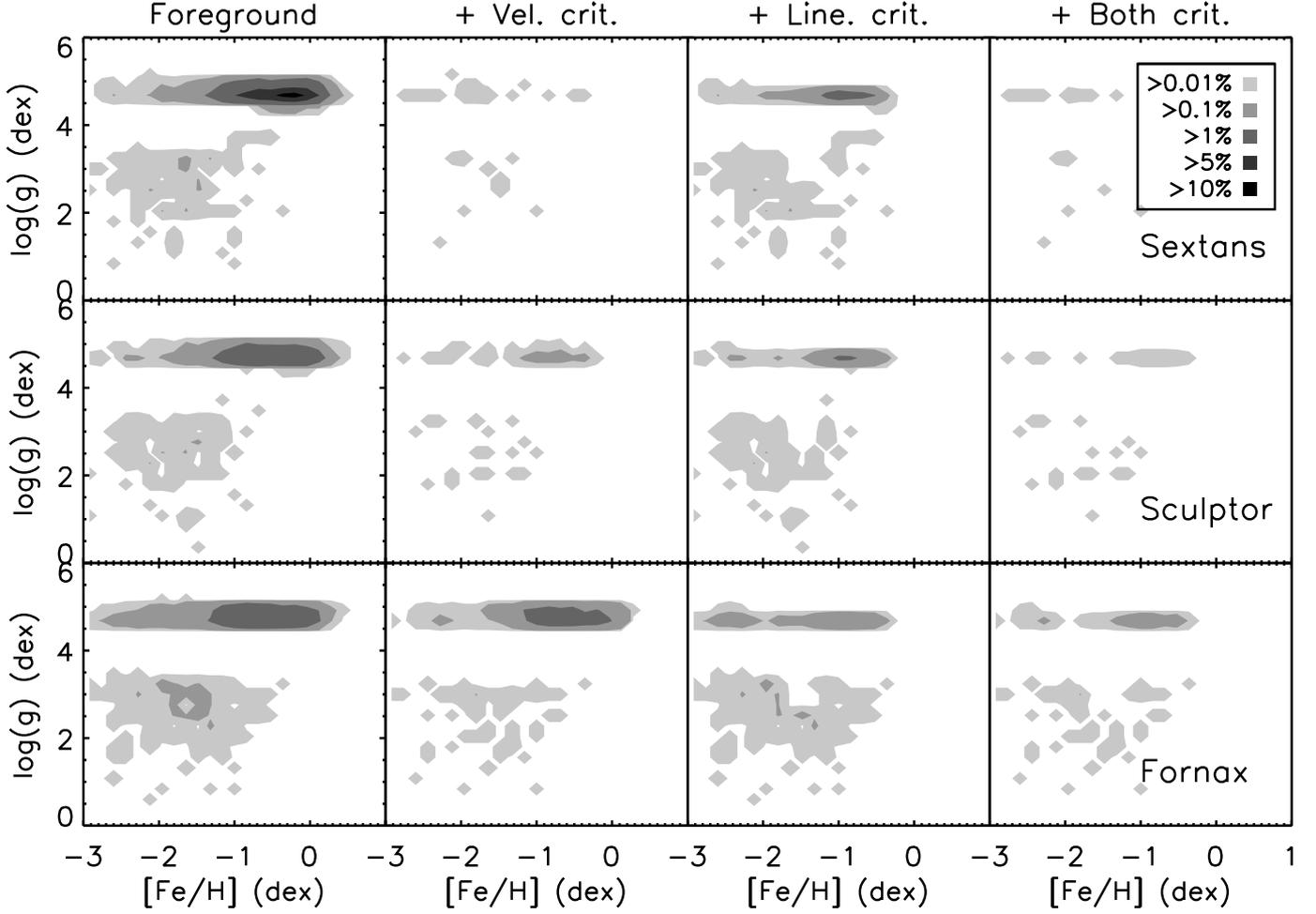}
\caption{Left panels: [Fe/H] and log(g) for all foreground stars drawn from the
  Besancon model towards the direction of the dwarf spheroidal
  galaxies Sextans (top row), Sculptor (middle row) and Fornax (bottom row). Contours are shown on a 25 by
  25 grid and represent percentages of 0.001, 0.01, 0.1 and 0.5 the
  total number of stars. The rest of the panels show subsequently the distribution of foreground stars left after applying a velocity criterion of 3$\sigma$ around the systemic velocity of the galaxy, a line criterion as described in the text on the basis of the EWs of the CaT and \mgi lines, and both criteria combined. Percentage levels are the same in all panels and always relative to the total number of foreground stars towards that galaxy.  \label{fig:fg_loggfeh}}
\end{figure*}

\begin{figure}[!ht]
\includegraphics[width=\linewidth]{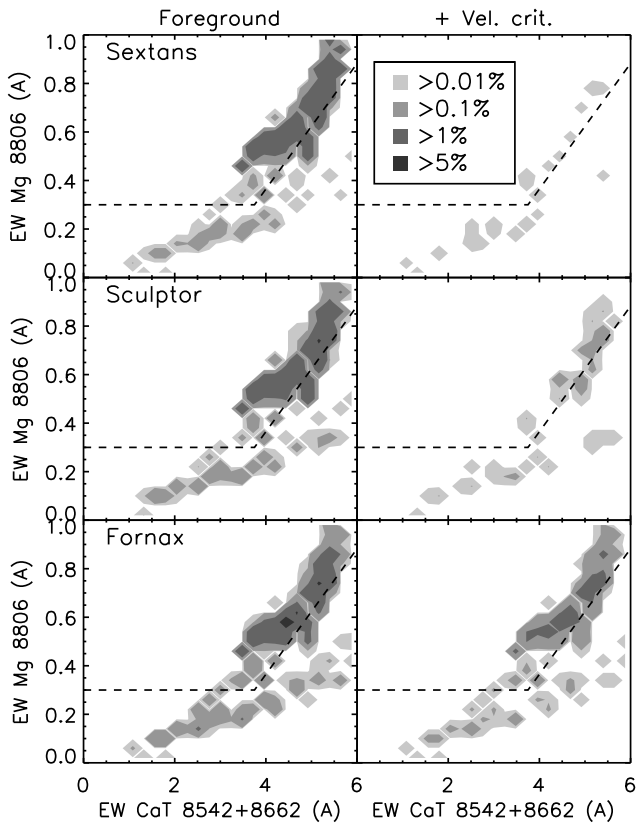}
\caption{\mgi EW and CaT $\Sigma {\rm W}$ for the foreground stars drawn from the Besancon model towards the direction of the dwarf spheroidal 
galaxies Sextans (top row), Sculptor (middle row) and Fornax (bottom row). The left panels show all the foreground stars, while the right panels the distribution of foreground stars left after applying 
a velocity criterion of 3$\sigma$ around the systemic velocity of the galaxy. The dashed line indicates Eq.~\ref{eq:line}. \label{fig:fg_mgcatew}}
\end{figure}

\begin{figure}[]
\begin{center}
\includegraphics[width=\linewidth]{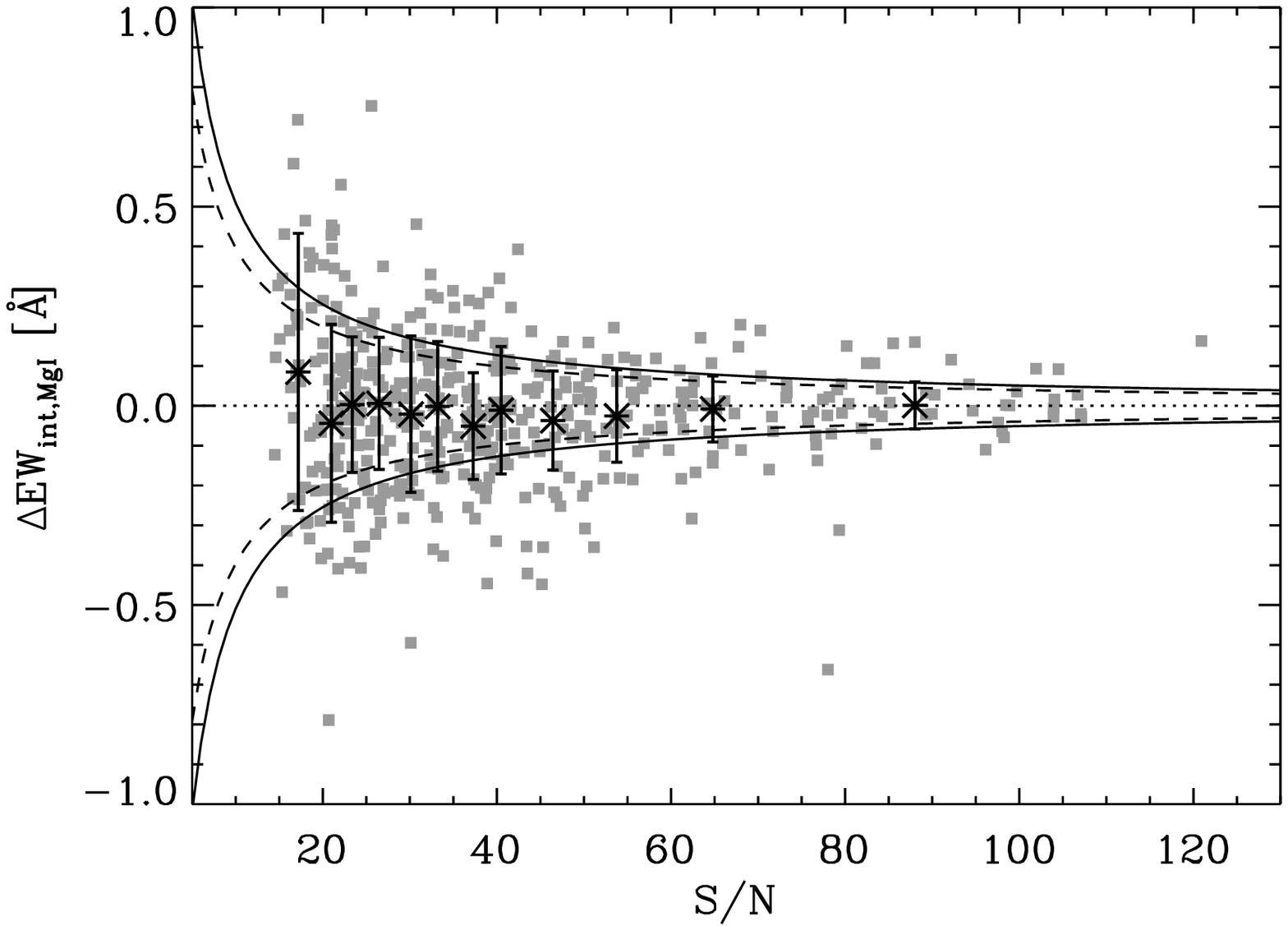}
\caption{Comparison between \mgi EW measurements for stars with double
  measurements for the Sextans, Sculptor and Fornax dSphs (59, 137, 289 stars, respectively). 
  The figure shows the distribution of \mgi EW differences
  from integrated flux as a function of S/N for the stars with S/N per
  \AA\,$\ge$ 10 and estimated error in velocity $\le$ 5 \kms for each
  measurement. The weighted mean in
  the \mgi EW difference ($\Delta \mathrm{EW_{int,MgI}}$), $rms$ dispersion and scaled median absolute
  deviation from the median are $0.01 \pm 0.02$ \AA\,, $0.09\pm 0.03$
  \AA\,, $0.08$ \AA\, for Sextans, $-0.04 \pm 0.03$ \AA\,, $0.3\pm
  0.03$ \AA\,, $0.19$ \AA\, for Fornax, $-0.02 \pm 0.01$ \AA\,,
  $0.15\pm 0.02$ \AA\,, $0.15$ \AA\, for Sculptor.  Assuming a similar S/N for the individual
  measurements of each star, a fit to the m.a.d. of these differences as 
  a function of S/N yields $\sqrt{2} \times \sigma_{\rm EW} = \sqrt{2}
  \times 3.6/(S/N)$. The solid lines indicate the 1-$\sigma$ 
  region for the error in the difference of EW; the dashed lines show the relation used in 
  \citet{batt11}. The asterisks indicate the median $\Delta \mathrm{EW_{int,MgI}}$ per S/N bin, 
  with the scaled m.a.d. shown as error-bar.} \label{fig:dew}
\end{center}
\end{figure}

\section{The foreground contamination}\label{sec:foreg}

\subsection{Fornax, Sculptor and Sextans}

As a test case of different mixtures of MW interlopers we look at the expected contamination along three different line-of-sights, specifically to the  Sextans, Sculptor and Fornax dSphs. 

To study the likely foreground contamination towards these objects, we used the Besan\c{c}on model \citep{robin03}. All interloper stars were selected within a 4 deg$^2$ solid angle around these galaxies, at a distance between 0 and 200 kpc. As in Sect.~\ref{sec:lines}, we restrict the analysis to the color range 0.85 $<$V-I$<$ 2.1. We also make a selection in apparent magnitude, which would include giant stars above the HB at the distance of these ``classical'' MW dSphs. For this we use the distance moduli from the literature and information on the HB magnitudes (see Table~\ref{tab:sel}), and we adopt an absolute magnitude in $I$-band $M_I = -4$ for the tip of the RGB. This range in apparent magnitude will be populated also by a range of MW contaminants - mainly main sequence dwarf stars - with the appropriate combination of absolute magnitude and distance from the Sun. 

The leftmost panels of Figure \ref{fig:fg_loggfeh} show the foreground sample in the direction of these three dwarf galaxies (note that in all these panels the contour levels are not linear), while the number of contaminants is given in Table~\ref{tab:sel}. It is clear that the type of stars within the foreground sample will depend on the direction in the sky.  Sculptor, being the galaxy at highest latitude has the least foreground in general, but a relatively high fraction of its foreground are giant stars, difficult to weed out with our method. Although all these galaxies are at high(ish) latitudes, there still is a very significant contamination from the disk in all cases, in particular from the thick disk. The thin and thick disk contribute almost entirely dwarf stars (log(g) $>$ 4.3), while the fraction drops to about 50\% for the contribution of dwarf stars from the halo. As previously discussed, the figure also shows that the great majority ($>$90\%) of the contaminants are found at [Fe/H]$>-2$ dex, where the distinction between giants and dwarfs is easier. The stars with [Fe/H]$<-2$ dex mostly belong to the stellar halo, which is the least dominant population in all of these cases.

In previous work, the member stars of the dwarf galaxy were selected out from the contaminants on the basis of their line-of-sight velocities. The stars were for example required to have a line-of-sight velocity within 3$\sigma$ from the systemic velocity of the galaxy \citep[for the Fornax dwarf galaxy 2.5$\sigma$ was used, because of its systemic velocity is closer to the velocity of the Galactic disk][]{batt06}. Using the simulated catalogs of MW contaminants we find that in all cases, as expected, some foreground stars will remain which have similar velocities as the dwarf galaxy stars, as is shown in the second column in Figure \ref{fig:fg_loggfeh}. The fraction though changes significantly from galaxy to galaxy. For Sculptor and Sextans the fraction of interlopers retained by the velocity criterion is low, as listed in Table \ref{tab:sel}, but the cleaning by velocity is much less effective in the Fornax dwarf galaxy, where 44\% of the interloper stars have a velocity within 3-$\sigma$ of the systemic velocity of Fornax (35\% if 2.5-$\sigma$ is used). In comparison, for Sextans this is just 0.9\% within 3-$\sigma$. 

In the third column of Figure~\ref{fig:fg_loggfeh}, we weed out contaminants according to the ``line criterion'', i.e. for which the closest synthetic spectrum in physical properties has a ${\rm EW}_{\rm Mg}/\Sigma {\rm W}_{\rm CaT}$ ratio which falls above the dividing line in Figure \ref{iso}. To derive the information on the line criterion all interloper stars from the Besan\c{c}on file were linked to the closest model within the grid of synthetic models (described in Section \ref{sec:lines}) in [Fe/H], log(g) and temperature space and the EWs of the CaT and \mgi lines of the closest model are used as their measurements. In all three galaxies it can clearly be seen that although a lot of contaminants are selected out (only 12 - 17\% remains), the foreground consisting of giants remains (see also Figure~\ref{fig:fg_mgcatew} for the distribution of contaminants in the EW$_{\rm Mg}$ vs $\Sigma {\rm W}_{\rm CaT}$ plane before and after velocity selection). The line criterion is less effective than the velocity criterion for Sextans and Sculptor. From Table~\ref{tab:sel} however is clear that for Fornax, that has a systemic velocity much closer to the one of MW thin disk with respect to the other two dSphs, the line criterion performs much better than the velocity criterion on its own. 

The best results in all cases are obtained when the velocity and the line criterion are used in addition to each other, as they are completely complementary in the sense that they will select out different sets of foreground stars. The line criterion will be selecting especially on the physical properties (mostly gravity) of the star, the velocity criterion just on its dynamics within the galaxy. Especially metal-rich dwarf foreground stars which have similar velocities and will be retained within the sample using a velocity cut, will be very easily separated by the method using absorption lines. In the last column of Figure \ref{fig:fg_loggfeh} we show the result of a combination of both methods. In all galaxies a very large fraction of the interlopers can be weeded out and only a few percent of the interlopers still remain.

An additional value of complementary use of both a velocity and line criterion is given in the last column of Table \ref{tab:sel}, where we relax the velocity criterion allowing for a selection of member stars within a larger range of velocities, e.g. 4-$\sigma$, and find that this does not cause a large increase in the number of interlopers. This can prove extremely useful, since one of the concerns of imposing strict velocity criteria is the possible biases introduced by the exclusion of member stars with more deviant velocities from the mean, for example when investigating the mass content of the galaxy from the velocity distribution of its stars.

\subsection{Predictions for the use of the method for other Milky Way dwarfs, including the ultra-faints} \label{sec:others}

We now extend the previous analysis to the other early type MW satellites. In the lower part of Table \ref{tab:sel} (below the horizontal line) we summarize the relevant information for all these galaxies for which dynamical analyses have been carried out, including the UFDs \citep{stri08}. For each of these galaxies we have extracted a catalog of foreground stars from the Besan\c{c}on models and analyzed it in the exact same way as described for the classical satellites Sextans, Sculptor and Fornax. In the various galaxies spectroscopic samples extend down to different depths, and most UFDs have been surveyed in observations also below the HB to increase the number of targets due to their sparsely populated RGB. For uniformity with the previous analysis we consider here only the region above the HB; note though that this method will be effective at the metallicities expected for stars in UFDs also down to approximately one magnitude below the HB, corresponding to the sub-giant branch boxes in Fig.~3 (see also the 
Appendix for an exploration of the method performance to fainter magnitudes 
and bluer colors).

As it can be seen from Table~\ref{tab:sel}, in all galaxies the line criterion can weed out approximately 80-90\% of the interlopers. The efficiency of a velocity criterion on the other hand, is much more dependent on the particular galaxy and its systemic velocity. In most cases either the velocity criterion is the most effective or performs rather comparably to the line criterion.  However, for Canes Venatici I,  Ursa Major I and Willman~1, the line criterion outperforms a velocity criterion. The difference is very significant for the latter two objects. For the three galaxies Hercules, Leo II and Leo IV, adding an extra criterion which is gravity-dependent will also pay off significantly. 

For all galaxies studied, a combination of both criteria would lead to an excellent cleaning of the samples, leaving only up to a few percent of the foreground contamination. This is particularly important for such intrinsically faint small galaxies, that have a sparsely populated RGB, where the samples of targets are bound to be small. In such cases, even if the number of contaminants is low, it can still represent a large fraction of the overall sample of observed stars. For example, let us take as a reference a faint system such as UMa~I, that has a luminosity in \mv\ band of $1.4 \pm 0.4 \times 10^4$ \sL \citep{mart08}, and calculate what would be the expected number of RGB stars with respect to MW interlopers for such a system. Using the publicly available synthetic color-magnitude diagram code provided by the {\it BaSTI} team\footnote{http:\/\/albione.oa-teramo.inaf.it\/.} and assuming that UMa~I can be approximated by a 13 Gyr old stellar population of [Fe/H]$=-2$ dex, about 20-25 stars are produced in the region of the RGB above the HB (the numbers are from 3 different random realizations) for a system of $\sim$ $2 \times 10^4$ \sL. For an exponential surface brightness profile, about 60\% of the light is produced within 1 exponential radius -- and therefore 60\% of the number of stars are contained within 1 exponential radius, assuming no mass segregation. Using 6.7 arcmin as the exponential radius of UMa~I \citep{mart08}, this would mean that approximately 12-15 RGB stars would be found over an area of $\sim$140 arcmin$^2$. Along the l.o.s. to UMa~I, the Besan\c{c}on model gives approximately 40 MW contaminants over such an area; using the velocity criterion on its own would result into retaining about 10 interlopers in the sample, i.e. in comparable number to the RGB stars of the dwarf over the same area. Using both a velocity and line criterion on the other hand would only leave 1-2 contaminants in the sample. 

Among the galaxies studied in this section, 
there are a few UFDs for which reaching about 2 mag down the HB appears to be particularly 
important in order to acquire statistically significant number of stars, such as for UMa~I, UMa~II, Leo~IV, Com, CVn~II and Wil~1. 
We explore the performance of the 
method to fainter magnitudes in the Appendix. We restrict the analysis to the most compelling cases, UMa~I, Leo~IV, Wil~1, since  
for UMa~II, Com and CVn~II the velocity criterion on its own is already very effective.

\section{The observed trend of Mg~I vs CaT EW} \label{sec:obs}

We subsequently explore the observed trend of \mgi EW versus $\Sigma {\rm W}_{\rm CaT}$ in actual data and how it compares to the predictions from the synthetic spectra, using the DART data-set for the 3 classical dSphs Sextans, Sculptor and Fornax. In this way we are able to empirically explore the use of the \mgi and CaT lines EWs for these dSphs with different properties - i.e. located along different line-of-sights, with different systemic velocities and whose stars cover different ranges of metallicities (see Table~\ref{tab:sel} for a summary). We used our intermediate resolution spectroscopic observations of these galaxies carried out at the VLT/FLAMES using the GIRAFFE spectrograph in Medusa mode with the LR8 grating, which covers the wavelength region 8206 -9400 \AA\, and therefore includes both the nIR CaT lines and the Mg~I line at 8806.8\AA\,. We refer the reader to the original papers for a general presentation of the data-set \citep{tols04, batt06, helm06, star10, batt11}. 

The data-reduction procedure adopted is described in detail in \citet{batt08}. Here we briefly remind the reader that the EW of the Mg~I line was derived by integrating the flux over 6 \AA\, around the central wavelength of the line as it was done for the synthetic spectra and in \citet{batt11}. For this line, which is smaller then the CaT lines, this EW estimator provides less noisy EW measurements with respect to a Gaussian fit to the line. 

The error in EW of the \mgi line was derived by using the stars with double measurements (see Fig.~\ref{fig:dew}). Assuming a similar S/N for the individual measurements of each star, we obtain a typical error in the EW of the \mgi line, $\sigma_{\rm EW}$, $0.06$ \AA\, for Sextans, $0.13$ \AA\, for Fornax, $0.11$ \AA\, for Sculptor as estimated from the scaled median absolute deviation from the median (m.a.d.) in the distribution of \mgi EW differences. We derive the trend of $\sigma_{\rm EW}$ with S/N by fitting the value of the scaled m.a.d. of the difference in EW$_{\rm Mg}$ as a function of S/N in bins containing at least 40 stars. Since the data-sets for Sextans, Sculptor and Fornax have been obtained using the same instrument, observing strategy and data-reduction, we join the data for the 3 galaxies. We obtain  $\Delta \mathrm{EW_{\rm Mg}} = \sqrt{2} \times \sigma_{\rm EW} = \sqrt{2} \times 3.6/(S/N)$ which would correspond to an error on the individual measurements of $\sigma_{\rm EW} = 3.6/(S/N)$. Figure~\ref{fig:dew} shows how this relation differs from the one used in \citet{batt11}, where only Sextans was analyzed.

\begin{figure*}[!ht]
\begin{center}
\includegraphics[width=0.32\linewidth]{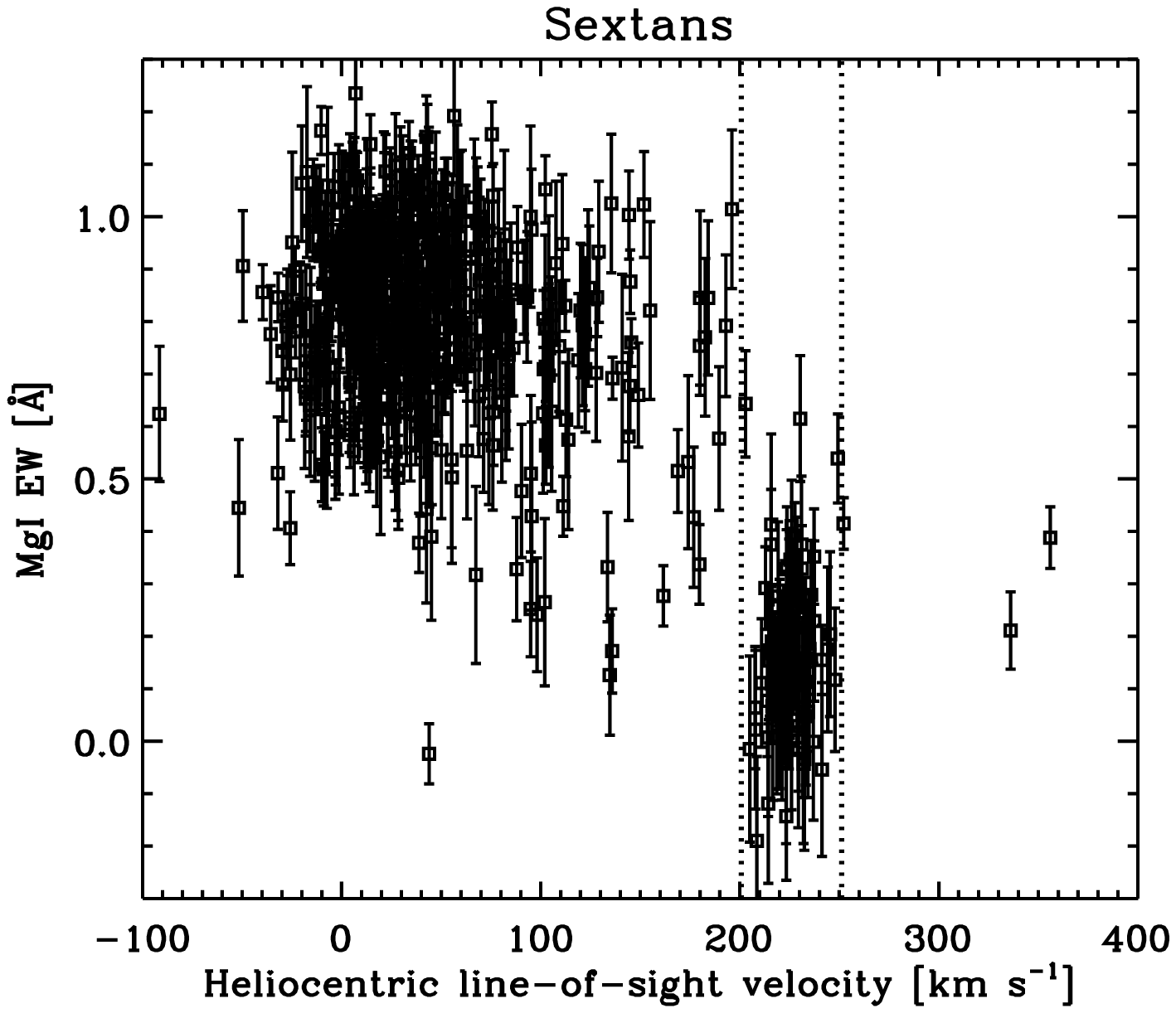}
\includegraphics[width=0.32\linewidth]{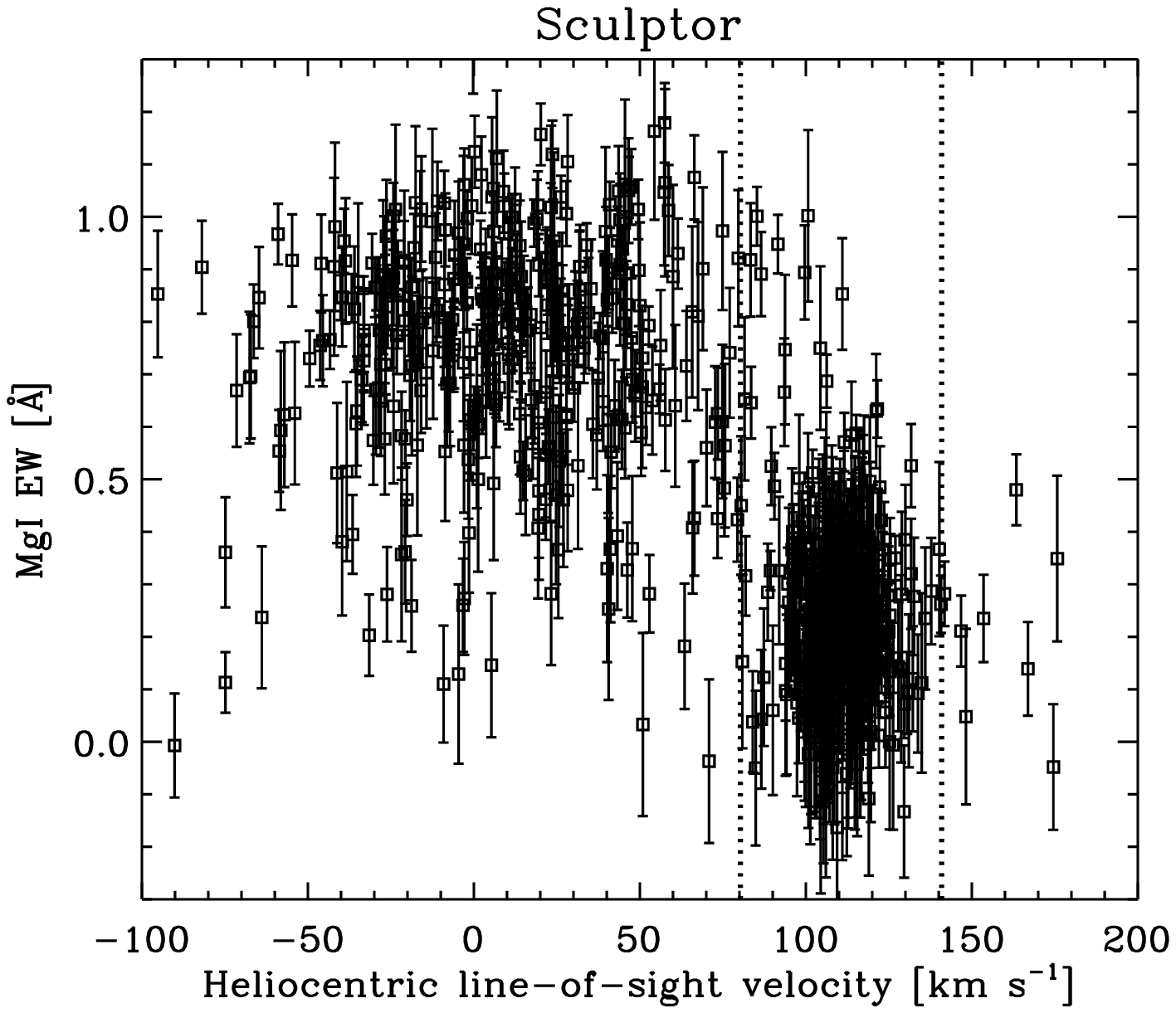}
\includegraphics[width=0.32\linewidth]{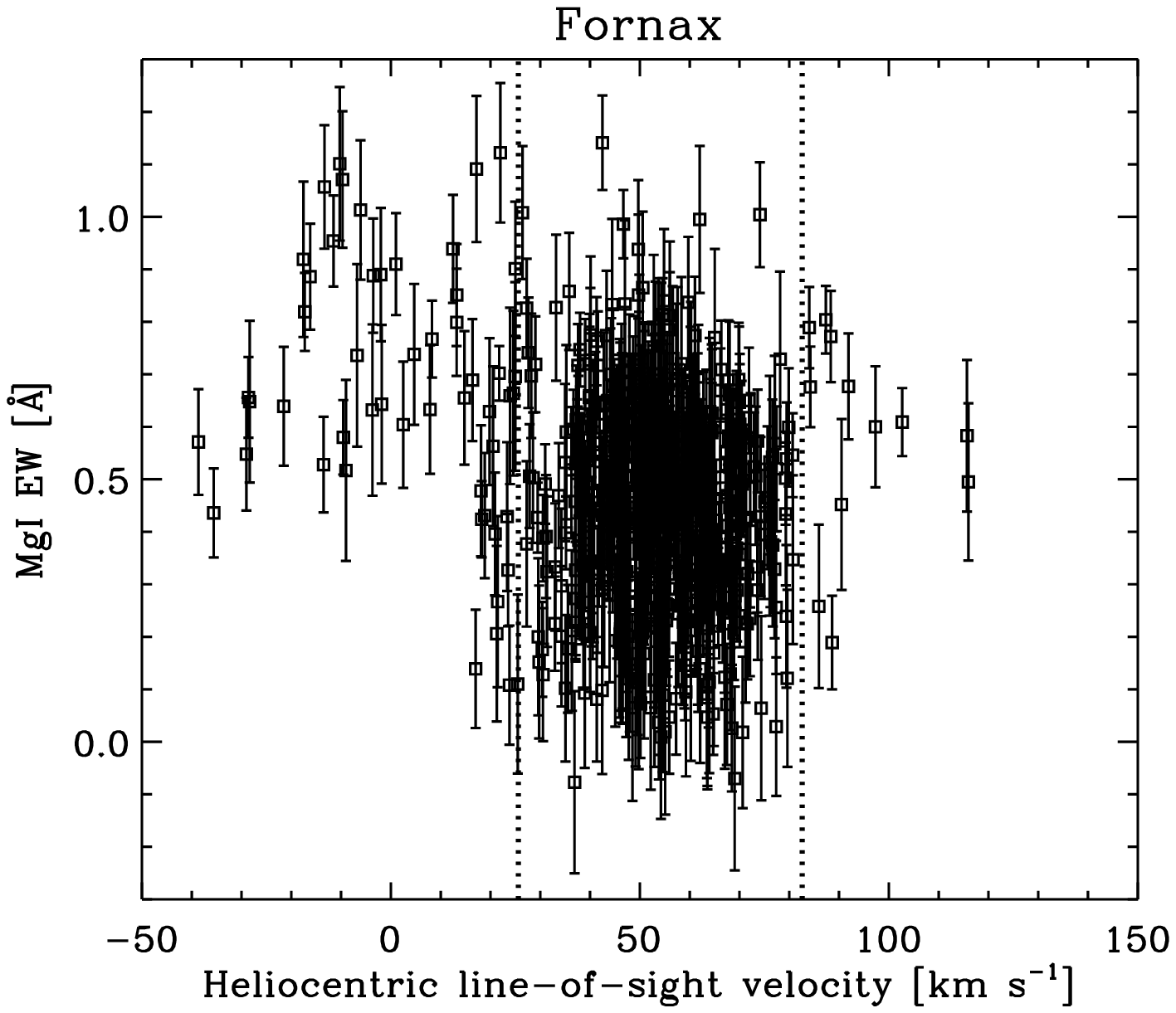}
\caption{\mgi EW versus line-of-sight velocity in the heliocentric
  system (squares with error-bars) for the Sextans, Sculptor and
  Fornax dSphs (from left to right). The vertical dotted lines show the region of kinematic membership 
  adopted in previous studies (i.e. with velocities within $\pm 3 \sigma$
  from the systemic velocity of the dSph for Sextans and Sculptor, and $\pm 2.5 \sigma$ 
  for Fornax). } \label{fig:mg_vel}
\end{center}
\end{figure*}

In Figure~\ref{fig:mg_vel} we show the behavior of the \mgi EW as a function of the line-of-sight velocity in the heliocentric system for the 3 dSphs. Here only spectra of higher quality (with S/N/\AA\, $>$ 20, velocity errors $<$ 5 \kms and for which the difference between the \mgi EW estimated from the flux integration and the Gaussian fit is within 5 times the scaled m.a.d. of the distribution of \mgi $|{EW_{\rm int} - EW_{\rm Gau}}|$) are shown. This has the purpose of allowing us to analyze the behavior of the \mgi line without being affected considerably by the noise in the measurements. The dotted lines indicate the region of kinematic membership adopted in previous studies ($\pm$3$\sigma$ from the systemic velocity of the galaxy for Sculptor and Sextans, and $\pm$2.5$\sigma$ for Fornax\footnote{This stricter kinematic criterion was used in previous work to reduce the foreground contamination given the low systemic velocity of Fornax}). In previous works stars within these velocity regions were considered as RGB stars probable members of the dSph. 

As the figure shows, the majority of what are clearly contaminant stars even by eye (i.e. those stars with line-of-sight velocity well outside the region of kinematic membership) can be weeded out on the basis of a kinematic selection. However, from the figure it is also clear that the velocity distribution of contaminants extends in, and also beyond, the region of kinematic membership. Obviously, when applying a velocity criterion on its own, there will be contaminants with velocities consistent with membership, and therefore classified as members. This is especially clear for the case of the Fornax dSph, whose low systemic velocity falls well within the velocity distribution of MW foreground stars. Concerning the distribution in \mgi EW, it can be seen that in all 3 cases the majority of contaminants show a rather distinct distribution in \mgi EW with respect to the bulk of probable dSphs members, i.e. they predominantly occupy the locus of \mgi EW $>$ 0.5 \AA\,. However a few differences are present among the galaxies, both in the distribution of probable kinematic members and non-members in terms of their \mgi EW. 

In the Sextans dSph 99\% of the stars with velocities within the region of kinematic membership have \mgi EW $<$ 0.5 \AA\,, which says that it is very unlikely that RGB stars - genuine members - will be found with \mgi EW $>$ 0.5 \AA\,. Therefore an hard-cut on the value of the \mgi EW as applied by B11 appears a good solution to weed out as many contaminants as possible with line-of-sight velocities consistent with kinematic membership, and at the same time retain in the sample the giants that are members of this dSph.

The same hard-cut though cannot be applied unchanged to the Sculptor and Fornax dSphs. In the case of Sculptor, it is clear from Fig.~\ref{fig:mg_vel} that  a larger fraction of stars well outside the region of kinematic membership have \mgi EWs $<$ 0.5 \AA\, with respect to Sextans. This is because of the different mix of MW stars along the line-of-sight to this galaxy (see Sect~\ref{sec:foreg} and Figure~\ref{fig:fg_mgcatew}). Furthermore, about 5\% of kinematic members extend to  \mgi EW $>$ 0.5 \AA\,, so that blindly imposing the same cut-off as Sextans could result in discarding a considerable fraction of genuine members of the dSph. In the case of Fornax, all but one of the kinematic non-members have \mgi EW $>$ 0.5 \AA\,, but also the kinematic members extend to large values of the \mgi EW, ranging from $\sim$0-1 \AA\,. Therefore applying the same cut-off in \mgi EW would allow to remove all contaminants, at the price of selecting out about 40\% of the stars with velocities consistent with membership! 

\begin{figure*}
\begin{center}
\includegraphics[width=0.32\linewidth]{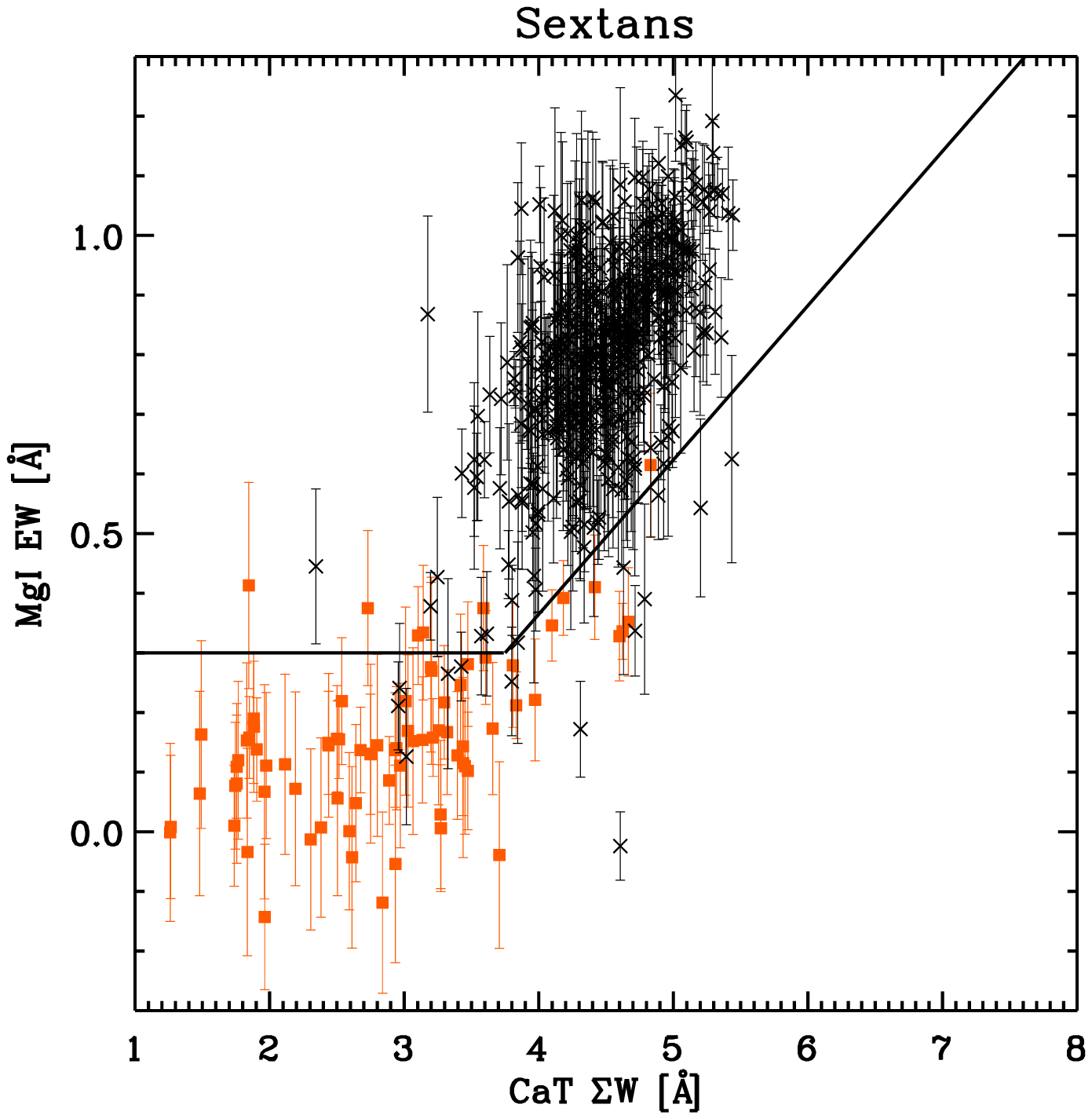}
\includegraphics[width=0.32\linewidth]{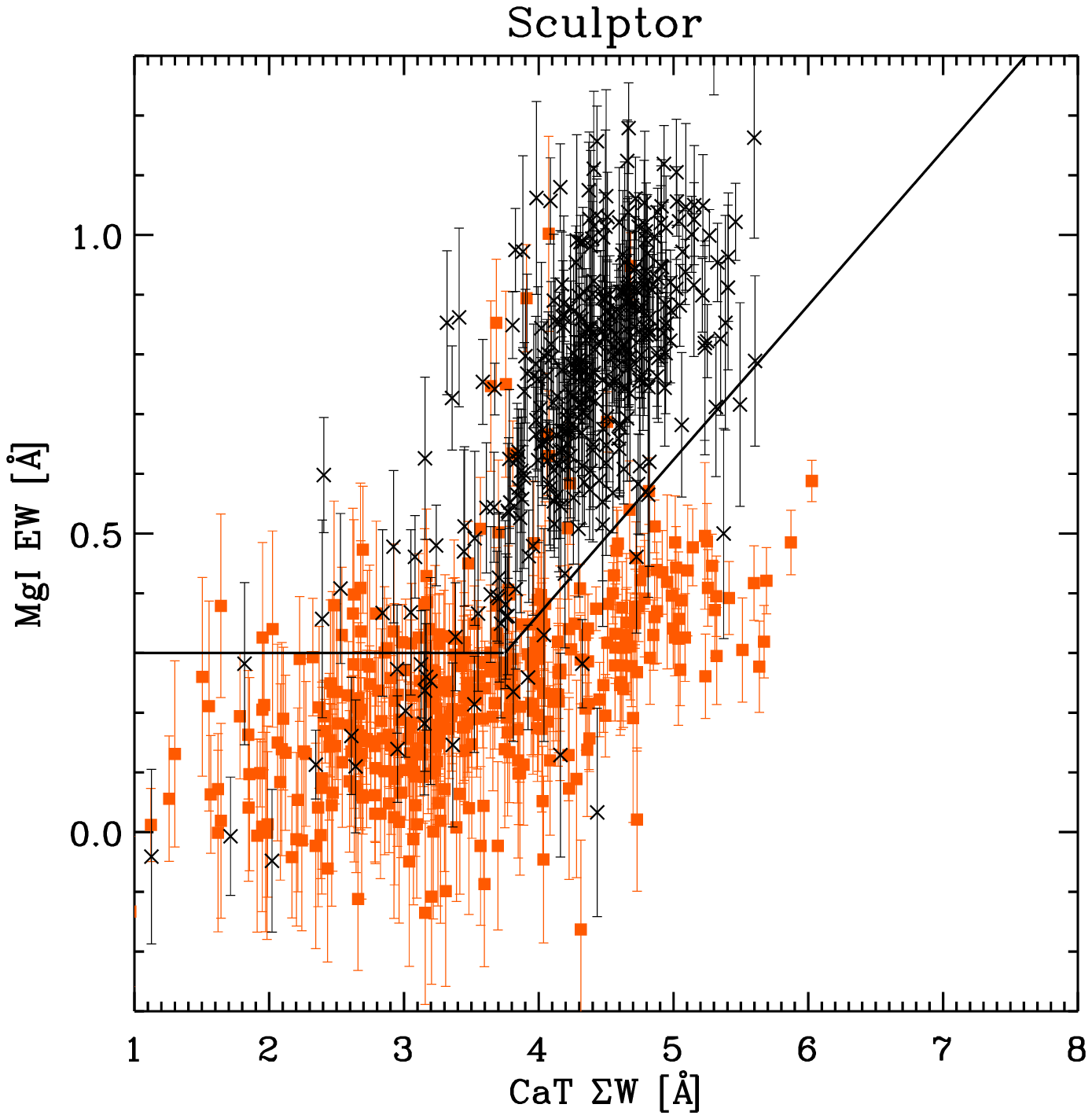}
\includegraphics[width=0.32\linewidth]{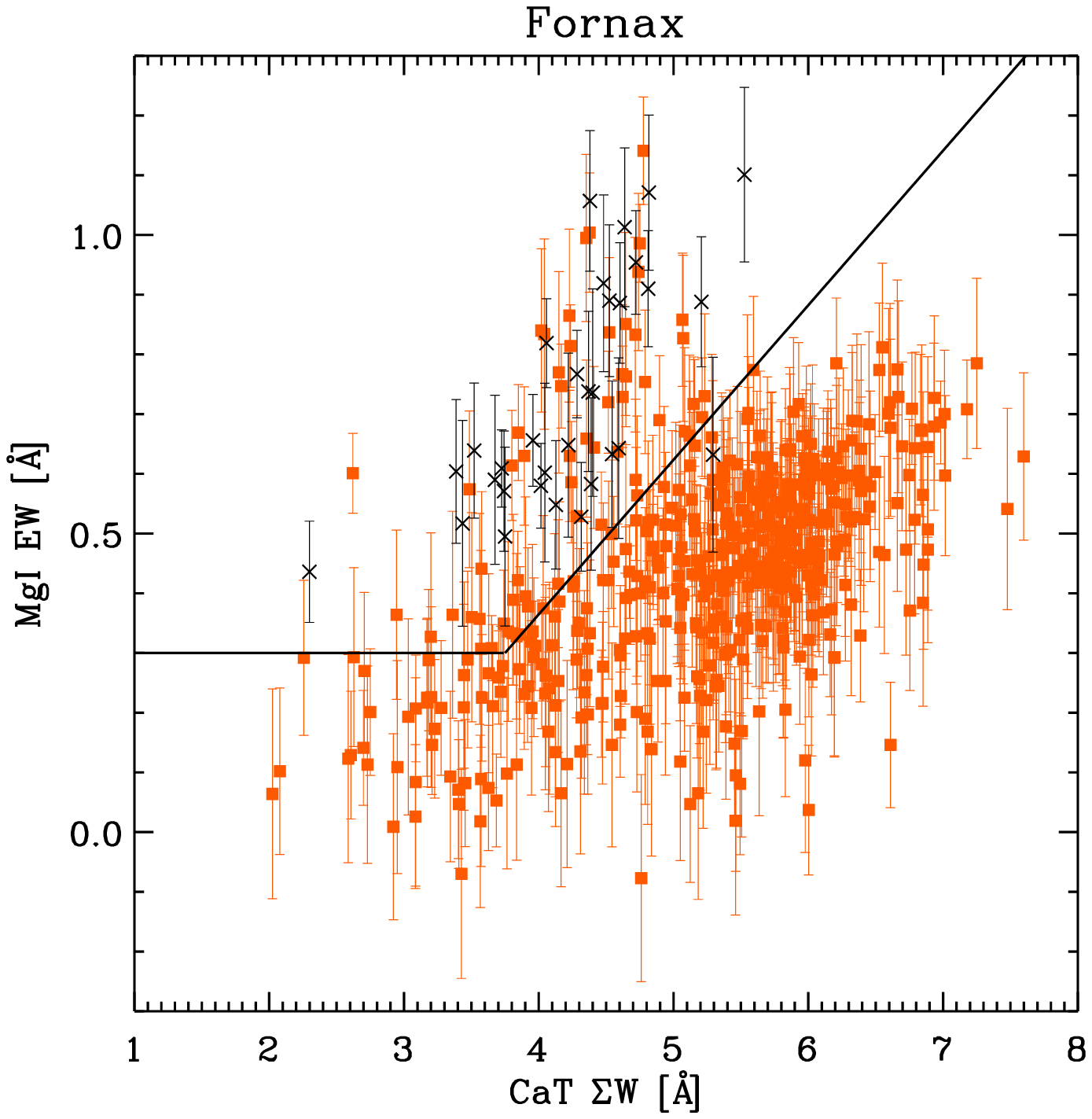}
\caption{\mgi EW versus $\Sigma  {\rm W}_{\rm CaT}$ for the Sextans, Sculptor and
  Fornax dSphs.  The filled orange squares with error-bars show the
  stars which are highly probable members (i.e. with velocities within
  $\pm 2 \sigma$ from the systemic velocity of the dSphs) and the
  black crosses the stars which are highly likely to be MW
  contaminants (i.e. with velocities at least $4 \sigma$ away from the
  systemic velocity of the dSphs). The solid line indicates the theoretical relation 
  derived in Sect.~\ref{sec:lines}. Note that because of the measurement errors the distribution of 
  highly probable members on this plane is inflated and in some cases goes over the separation line between 
  dwarf and giant stars.} \label{fig:mg_sumew}
\end{center}
\end{figure*}

The reason for this is clearly seen in Fig.~\ref{fig:mg_sumew} where we plot the distribution in the CaT $\Sigma W$- \mgi EW plane for highly probable members of the dwarf and highly likely non-members. Since these are real data, and we do not know a priori what is a contaminant and what a genuine member to the dSph, to increase the probability of selecting members and non-members and make this distinction easier using only l.o.s. velocities, we plot the stars within 2-$\sigma$ from the systemic as members, and the stars beyond 4-$\sigma$ from the systemic as non-members. We emphasize that 
here we are interested in the behaviour of the bulk of stars in both categories. For 
the likely members, a correlation is present between the $\Sigma {\rm W}_{\rm CaT}$ and the \mgi EW, so that stars with larger $\Sigma {\rm W}_{\rm CaT}$ display larger \mgi EW. At $\Sigma {\rm W}_{\rm CaT} \gtrsim$ 6 \AA\,, the typical \mgi EW is $>$0.5 \AA\,, and in Fornax, which is more metal-rich than Sextans and Sculptor, most of the stars in our data-set occupy the region $\Sigma  {\rm W}_{\rm CaT} \gtrsim$ 6 \AA\,. The correlation in \mgi EW vs $\Sigma  {\rm W}_{\rm CaT}$ observed in the data is consistent with the the synthetic spectra analysis of Sect.~\ref{sec:lines} and appears similar for the different dSphs at similar values of $\Sigma W$. This confirms the result of Sect.~\ref{sec:lines} for which a general criterion can be applied in \mgi EW vs $\Sigma  {\rm W}_{\rm CaT}$ to separate contaminants and probable members of the dSphs. Obviously in the case of observations the errors in the measurement of the EWs of the lines need also to be taken into consideration when comparing the location of the star in the \mgi vs $\Sigma  {\rm W}_{\rm CaT}$ plane with the line separating dwarfs and giants derived from the models.

\subsection{Applying the line criterion to data} \label{sec:clean}

In the previous sections we have proven the applicability and validity of the method; it is 
therefore interesting to see how the application of the line criterion to
available data-sets, e.g. DART, would perform. Here when referring to ``contaminants'' we 
refer to stars classified as such on the basis of the line criterion.

As visual inspection of Figure~\ref{fig:mg_sumew} already suggests, a significant number of stars which have l.o.s. velocities consistent with being highly probably members of the Fornax dSphs are actually found above the dividing line between dwarf and giant stars derived in Sect.~\ref{sec:lines}, and therefore are most likely MW interlopers that were misclassified because went unrecognized even by a strict velocity selection. A smaller number of such objects is also present in Sextans and Sculptor.

We apply the method to those stars that pass our usual selection criteria (i.e. S/N$>$10/\AA\,, velocity errors $<$ 5 \kms and CaT $|{\rm EW_{int}-EW_{gau}}| < 2$ \AA\,)\footnote{Since part of the targets for Sculptor were chosen from preliminary photometry, a handful of them turned out to have colors and magnitudes well outside the overall selection box for RGB stars from the definitive photometry. We exclude these objects by restricting ourselves to \vi $>$ 0.5 and V $>$ 16.5.}. In this work we account for the errors in \mgi EW in a simple way, i.e. we consider as ``giants'' those stars with ${\rm EW}_{\mgi}$ lower or within 1$\sigma_{\rm EW}$ from the \mgi EW obtained by applying Eq.~\ref{eq:line} (see dividing line in Figure~\ref{iso}). We define as ``contaminants'' those stars with \mgi EW more than 1$\sigma_{\rm EW}$ larger than the value obtained by applying Eq.~\ref{eq:line}. A more detailed approach, which would for example take into account the combined probability of membership from the velocity and \mgi EW distribution in a statistical way, is outside the scope of this paper; we refer the reader to \citet{walk09a} and \citet{mart11} for works dealing with the issue of membership with comprehensive statistical analyses.

The Sextans dSphs was previously analyzed; here we just mention that by considering the error in \mgi line in the selection, only 3 stars are considered as contaminants, instead than the 6 stars rejected in B11. This has no influence on the results other than bringing back into the sample stars which had discrepant metallicities with respect to the body of the distribution, indication that perhaps this selection may be too generous. The rejected stars account for less than 2\% of the overall sample. This may seem surprising since Sextans is a very low surface brightness object, with a CMD heavily contaminated by MW stars, and therefore one may expect a large amount of contaminants. However, the systemic velocity of the object, far from the one of the MW disks, appears to allow for a good cleaning of the sample just on the basis of the l.o.s. velocity. 

Among the stars within the region of kinematic membership, Fornax and Sculptor both have a percentage of $\sim$5\% of stars classified as contaminants according to the line criterion. Note that, for Fornax, kinematic members were chosen from a 2.5$\sigma$ cut in our previous works; if we consider the kinematic membership region as the velocity range between $\pm$3$\sigma$ from the systemic velocity, the fraction of interlopers increases to 6.5\% of the total. These contaminant stars are found at all projected radii, but their relative contribution to the number of members becomes increasingly significant towards the outer parts, accounting for example for about 45\% and 35\% of the kinematic members at $R > $1.0 deg for Fornax and Sculptor, respectively. For comparison,  between 0.5 and 1.0 deg \citep[approx 0.4 and 0.8 nominal tidal radii, according to the value from][]{ih95} the fraction drops to 6.3\% in Sculptor. The fraction of contaminants found for these two objects are very similar to each other, even though Table~1 would have suggested that a much larger fraction would be found for Fornax. This is most likely due to a combination of factors such as the different area covered and the much larger surface brightness of Fornax with respect of Sculptor, which permits to pick out a larger number of member stars from the foreground. 

\begin{figure*}[!ht]
\begin{center}
\includegraphics[width=0.45\linewidth]{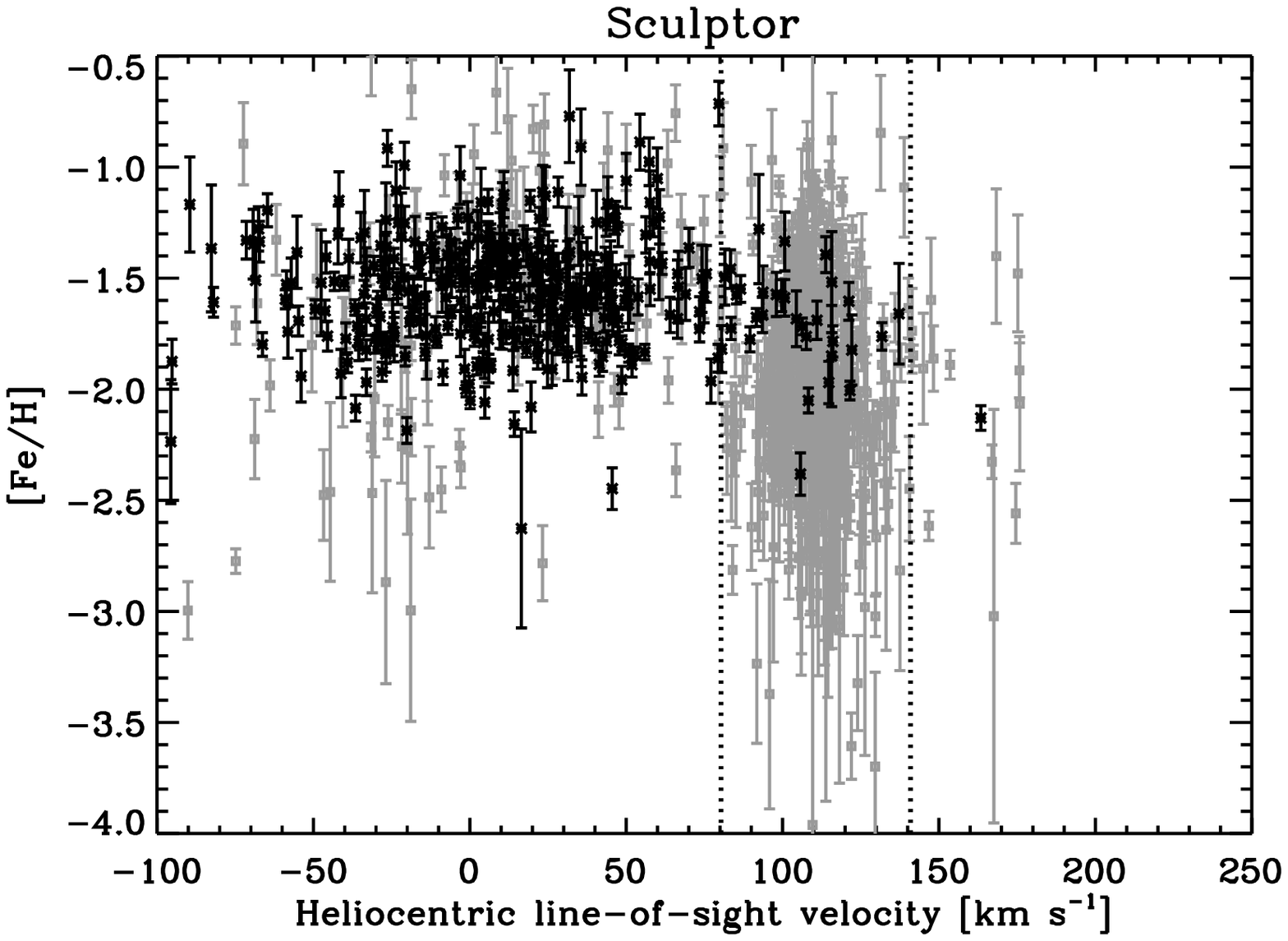}
\includegraphics[width=0.45\linewidth]{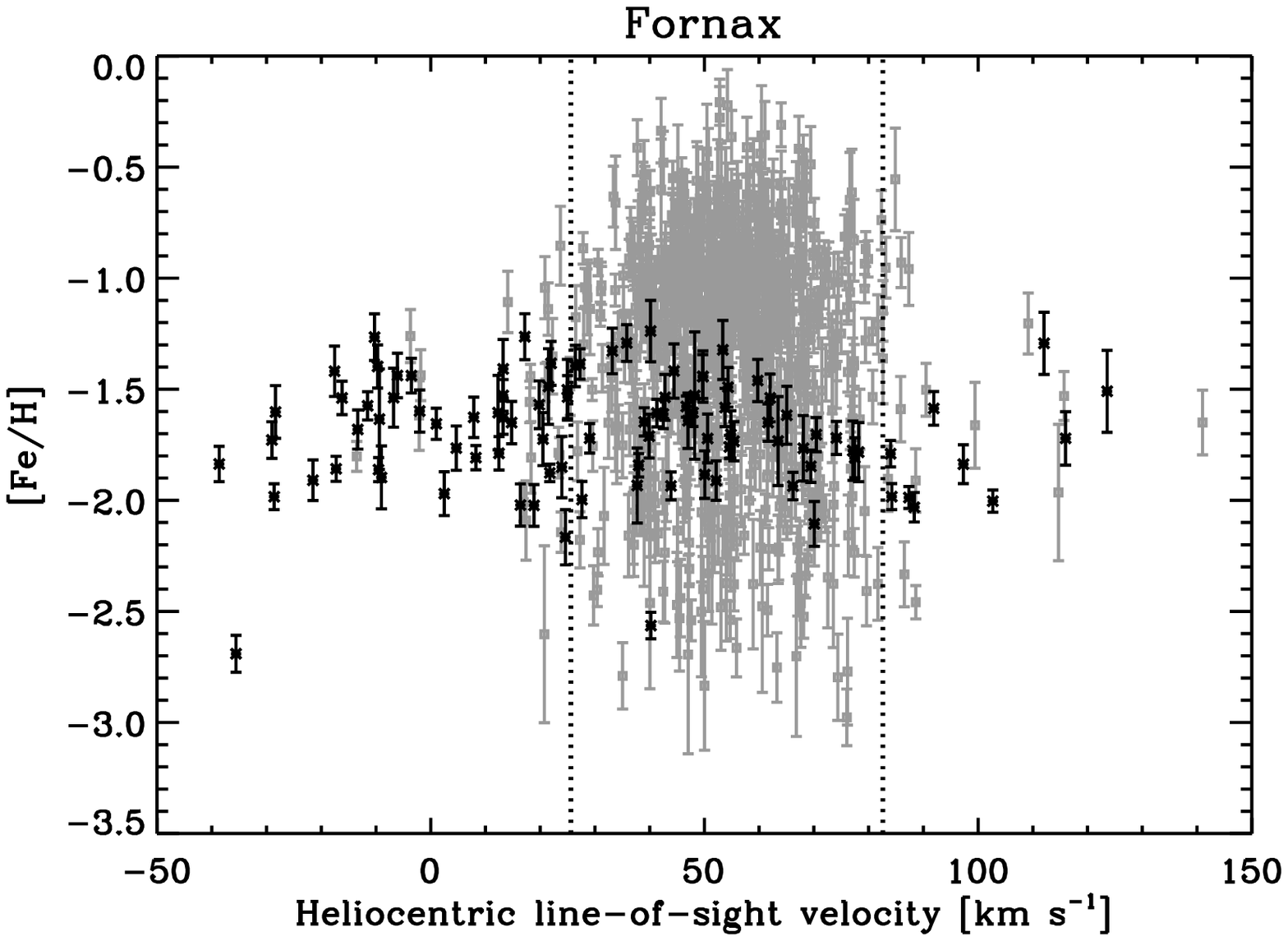}
\includegraphics[width=0.45\linewidth]{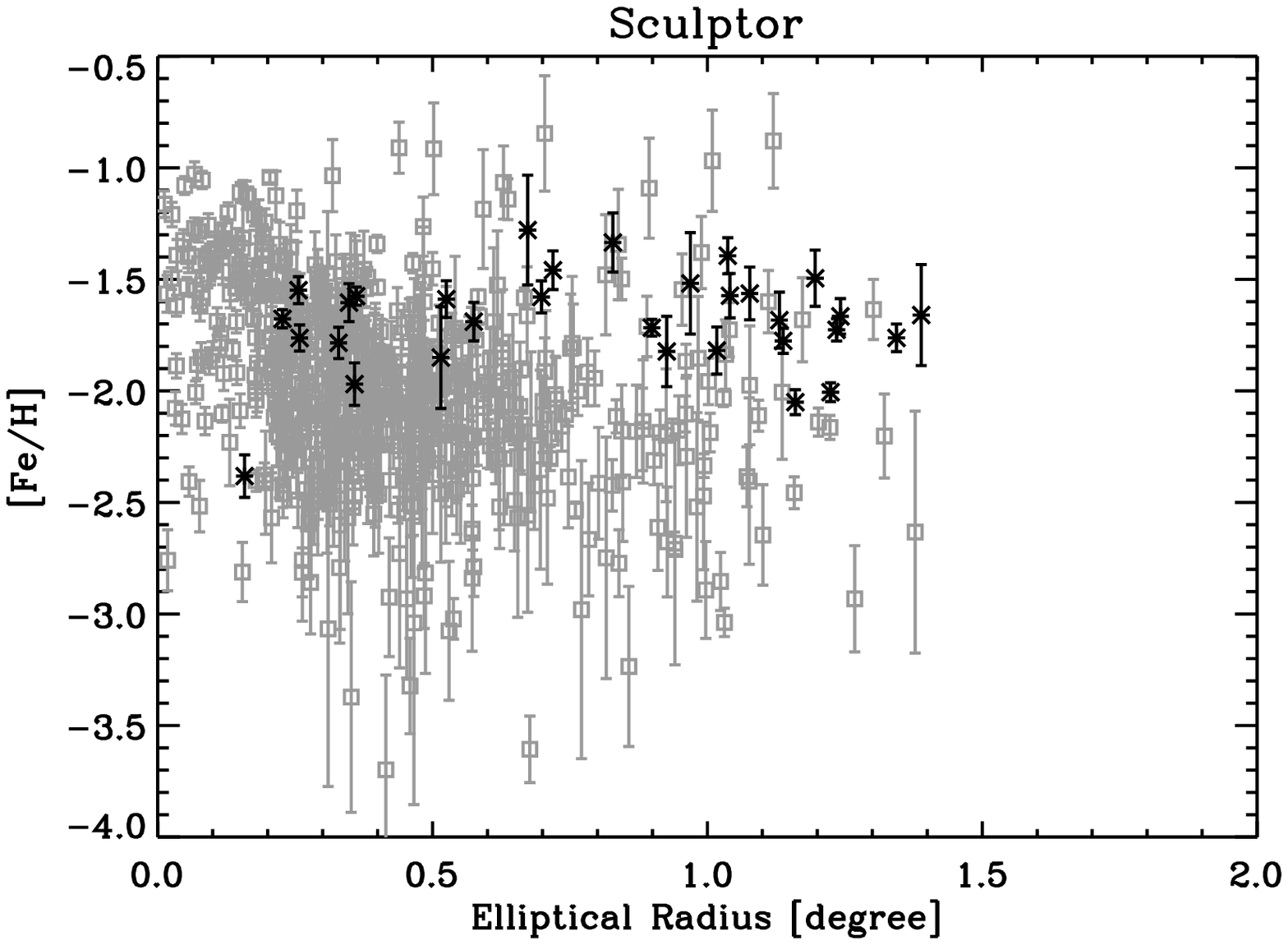}
\includegraphics[width=0.45\linewidth]{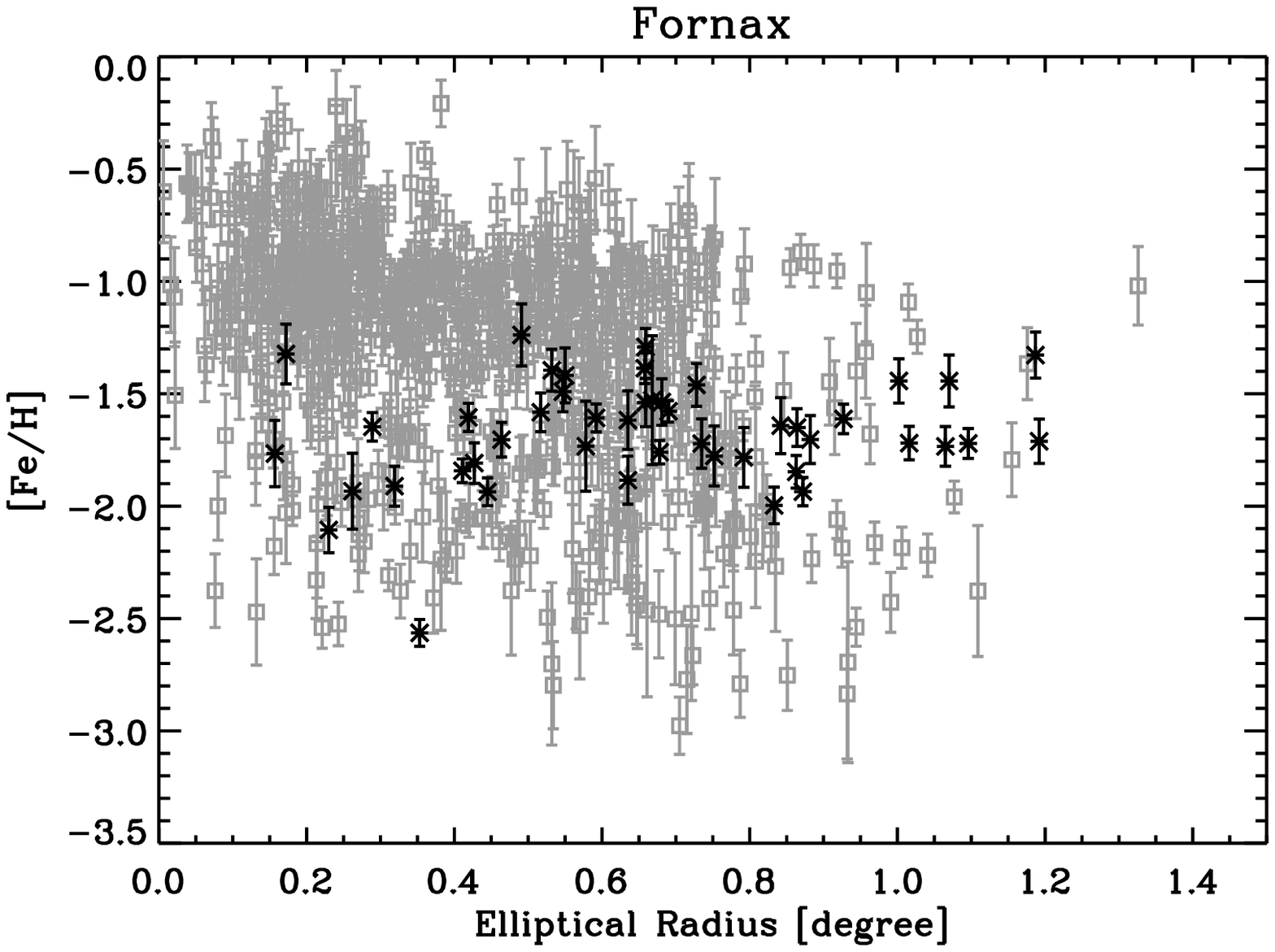}
\caption{Application of the line method to the Sculptor (left) and Fornax (right) dSphs, using the
 VLT/FLAMES DART data-set. Top panels: Distribution on the CaT [Fe/H] versus l.o.s. heliocentric velocity 
for the stars with S/N/\AA\, $>$10, error in velocity $<$ 5 \kms and CaT $|{\rm EW_{int}-EW_{gau}}| < 2$ \AA\,. 
The black squares with error bars show those stars whose \mgi EW was either lower or within 
1$\sigma_{\rm EW}$ from the \mgi EW given by Eq.~\ref{eq:line}, while cyan (grey in the black and white version) squares 
with error bars show those stars whose \mgi EW was more than 
1$\sigma_{\rm EW}$ larger than the \mgi EW given by Eq.~\ref{eq:line}. The vertical lines indicate the region of kinematic 
membership from our previous works. Bottom panel: CaT [Fe/H] versus elliptical radius for 
kinematic members (black: with \mgi EW within 1$\sigma_{\rm EW}$ from Eq.~\ref{eq:line}; cyan: with \mgi EW more than 
1$\sigma_{\rm EW}$ away from Eq.~\ref{eq:line}.} \label{fig:fevel}
\end{center}
\end{figure*}

Figure~\ref{fig:fevel} (top) shows the distribution of the analyzed stars for Sculptor and Fornax on the CaT [Fe/H] versus heliocentric l.o.s. velocity plane. The combination of these quantities can already provide information on the probability of membership of a star to the MW or to the dwarf galaxy, as on this plane the great majority of those stars which are clearly unbound to the dwarf galaxy (MW stars) has a different distribution than the velocity members (likely RGB stars of the dSphs). For example, it is clear that the region with  $-2 <$ [Fe/H]$< -1$ is expected to be the most contaminated one. The figure also shows that most of the stars excluded by the line criterion alone indeed have CaT [Fe/H] values in the range where most contaminants were expected, highlighting the fact that the method here presented, which makes no a priori selection in metallicity and velocity, is indeed able to find interlopers. 

We note that the CaT [Fe/H] relation is empirically calibrated for RGB stars (above the HB) and relies on the assumption that all the stars can be considered at the same distance. While this is a reasonable assumption for the stars belonging to the dSphs here analyzed, this is not the case for the bulk of MW interlopers, which are located over a range of distances - mostly within a few kpc from the Sun - and have markedly different gravities than RGB stars. Therefore, the CaT [Fe/H] values for the stars which are probable non-members are not necessarily indicative of their  
true metallicity.

Figure~\ref{fig:fevel} (top) also illustrates what would be the result of relaxing the kinematic criterion to 4$\sigma$ for the 2 galaxies\footnote{Here we use the velocity dispersion values from Table~1.}: the fraction of interlopers becomes about 8\% for Fornax and 6.5\% for Sculptor; and the new contaminants are found at velocities lower than the systemic, towards the velocities of the disk, as it would be expected. The bottom panels of Figure~\ref{fig:fevel} show the metallicity distributions as a function of projected radius resulting from applying the line criterion in addition to previous kinematic criteria: the conclusions from our previous works in terms of changes of metallicity with radius holds. Also the smaller velocity dispersion of metal-rich stars with respect to metal-poor stars is still present. For Sculptor it can be seen that several of the stars with [Fe/H]$>-1.7$ dex at R$>$0.6 deg may actually not belong to Sculptor, making the variation of metallicity properties with projected radius even clearer.

One can therefore conclude that the main properties of systems for which samples of several hundreds individual stars are available, such as the majority of classical dSphs, are robust. It is also clear that methods of the kind here proposed can potentially be very helpful for those analyses whose conclusions heavily rely on the properties of stars in the outer parts of these systems, where the ratio of unrecognized interlopers to member stars increases, as well for studies of several of the fainter dwarf galaxies such as the UFDs. 

\section{Discussion and conclusions} \label{sec:conc}

In this work we have addressed the problem of identifying MW stars in samples of kinematically selected members stars in Local Group galaxies. The presence of such interlopers may contaminate the derived properties of the target galaxy and this is a general issue in studies of resolved stellar populations of Local Group galaxies.

The method that we propose is mainly meant for spectroscopic samples in the CaT region whose targets are selected from broad band photometry to lie on the a region of the CMD covering the RGB of the target galaxy (above or approximately one magnitude below the HB). This is the case for the great majority of data for ``classical'' dSphs and more distant Local Group dwarf galaxies; for several UFDs a significant fraction of the targets lie in that magnitude range. We have explored the combined use of the \mgi EW at 8806.8 \AA\, and the nIR CaT $\Sigma {\rm W}$ as a way of distinguishing between contaminants and members on the basis of physical characteristics of the star, mainly gravity, since the targeted stars are RGB stars while the MW interlopers will mostly consist of dwarf stars. 

For this we have used synthetic spectra over a range of metallicity, temperature and gravity covering the expected age and metallicity range of stars in Local Group dwarf galaxies. We found that a relation between the \mgi EW and the CaT $\Sigma {\rm W}$ can be applied to distinguish between dwarf and giant stars above the HB for metallicities [Fe/H] $> -2$ dex and for (sub)giant stars below the HB for -2$\le$ [Fe/H] $\le$-1. Since at lower metallicities the main contribution of contaminants comes from the MW stellar halo, which is the least dominant component in terms of stellar mass (and therefore number of stars), only a negligible amount of interlopers is anyway expected below this metallicity.

The relation between the \mgi EW and the CaT $\Sigma$~W derived from the analysis of synthetic spectra applies well to the trend observed in actual data for a sample of 3 classical dSphs (Sextans, Fornax and Sculptor), whose stars cover different ranges of age and metallicity. The line-of-sight to the target object and the velocity of the center of mass of the galaxy (the systemic velocity) are important parameters because they determine the amount and mix of contaminants that will be present in the kinematically selected samples of stars. This means that there will be objects for which the l.o.s. velocity selection is already very efficient, while other ones that greatly benefit from an additional criterion such as the one here explored. We find that the addition of this method to the information from the l.o.s. velocity can be particularly beneficial for example for Canes Venatici I, Leo~II, Fornax, Hercules, Ursa Major I and Willman~1. For the latter four galaxies, the use of the \mgi EW method on its own is considerably more efficient than the selection in l.o.s. velocity. 

It should be noted that the contamination due to giant stars from the MW is not removed by this method, the proposed relation will retain as many giant stars as possible and throw away only dwarf contaminants. Another very important parameter in determining the ratio between contaminants and stars member to the dwarf is the ratio between the projected number density of MW and dwarf galaxy stars. Besides varying from object to object, this ratio changes within the object itself simply because of the declining surface density of stars in dwarf galaxies, so that weeding out interlopers will become particularly important in the outer parts of the systems, and for intrinsically faint objects.

This method has the advantage of being very easily implemented, as it relies on measuring EWs for relatively large lines such as the nIR CaT and \mgi. These lines are close in wavelength, so that several existing studies of MW dSphs covering this wavelength region could benefit from this extra information.

\begin{acknowledgements} 
The authors would like to thank Mike Irwin for useful conversations that have inspired this work, comments and for the use of his equivalent width extraction routine. We thank Eline Tolstoy, Amina Helmi and Vanessa Hill for useful comments. We also kindly acknowledge Bertrand Plez for making his linelists and Turbospectrum code available. ES gratefully acknowledges support through an NWO-VICI grant. We acknowledge the International Space Science Institute (ISSI) at Bern for their funding of the team 
``Defining the full life-cycle of dwarf galaxy evolution: the Local Universe as a template''. 
The research leading to these results has received funding from the European Union
Seventh Framework Programme (FP7/2007-2013)
under grant agreement number PIEF-GA-2010-274151. This work has made use of {\it BaSTI} web tools. 
The authors would like to thank the {\it BaSTI} team for making publicly available the synthetic colour-magnitude 
diagram web tool. GB is grateful to the Kapteyn Astronomical Institute for hospitality and financial support during part of this work. 
\end{acknowledgements}

\bibliographystyle{aa}
\bibliography{art_Mgline_v10}

\begin{thebibliography}{50}
\expandafter\ifx\csname natexlab\endcsname\relax\def\natexlab#1{#1}\fi

\bibitem[{{Ad{\'e}n} {et~al.}(2009a){Ad{\'e}n}, {Feltzing}, {Koch},
  {Wilkinson}, {Grebel}, {Lundstr{\"o}m}, {Gilmore}, {Zucker}, {Belokurov},
  {Evans}, \& {Faria}}]{aden09a}
{Ad{\'e}n}, D., {Feltzing}, S., {Koch}, A., {et~al.} 2009a, \aap, 506, 1147

\bibitem[{{Ad{\'e}n} {et~al.}(2009b){Ad{\'e}n}, {Wilkinson}, {Read},
  {Feltzing}, {Koch}, {Gilmore}, {Grebel}, \& {Lundstr{\"o}m}}]{aden09b}
{Ad{\'e}n}, D., {Wilkinson}, M.~I., {Read}, J.~I., {et~al.} 2009b, \apjl, 706,
  L150

\bibitem[{{Alvarez} \& {Plez}(1998)}]{alva98}
{Alvarez}, R. \& {Plez}, B. 1998, \aap, 330, 1109

\bibitem[{{Barklem} {et~al.}(2005){Barklem}, {Christlieb}, {Beers}, {Hill},
  {Bessell}, {Holmberg}, {Marsteller}, {Rossi}, {Zickgraf}, \&
  {Reimers}}]{bark05}
{Barklem}, P.~S., {Christlieb}, N., {Beers}, T.~C., {et~al.} 2005, \aap, 439,
  129

\bibitem[{{Battaglia} {et~al.}(2008{\natexlab{a}}){Battaglia}, {Helmi},
  {Tolstoy}, {Irwin}, {Hill}, \& {Jablonka}}]{batt08scl}
{Battaglia}, G., {Helmi}, A., {Tolstoy}, E., {et~al.} 2008{\natexlab{a}},
  \apjl, 681, L13

\bibitem[{{Battaglia} {et~al.}(2008{\natexlab{b}}){Battaglia}, {Irwin},
  {Tolstoy}, {Hill}, {Helmi}, {Letarte}, \& {Jablonka}}]{batt08}
{Battaglia}, G., {Irwin}, M., {Tolstoy}, E., {et~al.} 2008{\natexlab{b}},
  \mnras, 383, 183

\bibitem[{{Battaglia} {et~al.}(2011){Battaglia}, {Tolstoy}, {Helmi}, {Irwin},
  {Parisi}, {Hill}, \& {Jablonka}}]{batt11}
{Battaglia}, G., {Tolstoy}, E., {Helmi}, A., {et~al.} 2011, \mnras, 411, 1013

\bibitem[{{Battaglia} {et~al.}(2006){Battaglia}, {Tolstoy}, {Helmi}, {Irwin},
  {Letarte}, {Jablonka}, {Hill}, {Venn}, {Shetrone}, {Arimoto}, {Primas},
  {Kaufer}, {Francois}, {Szeifert}, {Abel}, \& {Sadakane}}]{batt06}
{Battaglia}, G., {Tolstoy}, E., {Helmi}, A., {et~al.} 2006, \aap, 459, 423

\bibitem[{{Belokurov} {et~al.}(2007){Belokurov}, {Zucker}, {Evans}, {Kleyna},
  {Koposov}, {Hodgkin}, {Irwin}, {Gilmore}, {Wilkinson}, {Fellhauer},
  {Bramich}, {Hewett}, {Vidrih}, {De Jong}, {Smith}, {Rix}, {Bell}, {Wyse},
  {Newberg}, {Mayeur}, {Yanny}, {Rockosi}, {Gnedin}, {Schneider}, {Beers},
  {Barentine}, {Brewington}, {Brinkmann}, {Harvanek}, {Kleinman}, {Krzesinski},
  {Long}, {Nitta}, \& {Snedden}}]{belo07}
{Belokurov}, V., {Zucker}, D.~B., {Evans}, N.~W., {et~al.} 2007, \apj, 654, 897

\bibitem[{{Catelan} \& {Cort{\'e}s}(2008)}]{cate08}
{Catelan}, M. \& {Cort{\'e}s}, C. 2008, \apjl, 676, L135

\bibitem[{{Coleman} {et~al.}(2005){Coleman}, {Da Costa}, {Bland-Hawthorn}, \&
  {Freeman}}]{cole05}
{Coleman}, M.~G., {Da Costa}, G.~S., {Bland-Hawthorn}, J., \& {Freeman}, K.~C.
  2005, \aj, 129, 1443

\bibitem[{{Demarque} {et~al.}(2004){Demarque}, {Woo}, {Kim}, \& {Yi}}]{dema04}
{Demarque}, P., {Woo}, J., {Kim}, Y., \& {Yi}, S.~K. 2004, \apjs, 155, 667

\bibitem[{{Faria} {et~al.}(2007){Faria}, {Feltzing}, {Lundstr{\"o}m},
  {Gilmore}, {Wahlgren}, {Ardeberg}, \& {Linde}}]{fari07}
{Faria}, D., {Feltzing}, S., {Lundstr{\"o}m}, I., {et~al.} 2007, \aap, 465, 357

\bibitem[{{Gilbert} {et~al.}(2006){Gilbert}, {Guhathakurta}, {Kalirai}, {Rich},
  {Majewski}, {Ostheimer}, {Reitzel}, {Cenarro}, {Cooper}, {Luine}, \&
  {Patterson}}]{gilb06}
{Gilbert}, K.~M., {Guhathakurta}, P., {Kalirai}, J.~S., {et~al.} 2006, \apj,
  652, 1188

\bibitem[{{Gilmore} {et~al.}(2007){Gilmore}, {Wilkinson}, {Wyse}, {Kleyna},
  {Koch}, {Evans}, \& {Grebel}}]{gilm07}
{Gilmore}, G., {Wilkinson}, M.~I., {Wyse}, R.~F.~G., {et~al.} 2007, \apj, 663,
  948

\bibitem[{{Gustafsson} {et~al.}(2008){Gustafsson}, {Edvardsson}, {Eriksson},
  {J{\o}rgensen}, {Nordlund}, \& {Plez}}]{gust08}
{Gustafsson}, B., {Edvardsson}, B., {Eriksson}, K., {et~al.} 2008, \aap, 486,
  951

\bibitem[{{Helmi} {et~al.}(2006){Helmi}, {Irwin}, {Tolstoy}, {Battaglia},
  {Hill}, {Jablonka}, {Venn}, {Shetrone}, {Letarte}, {Arimoto}, {Abel},
  {Francois}, {Kaufer}, {Primas}, {Sadakane}, \& {Szeifert}}]{helm06}
{Helmi}, A., {Irwin}, M.~J., {Tolstoy}, E., {et~al.} 2006, \apjl, 651, L121

\bibitem[{{Irwin} \& {Hatzidimitriou}(1995)}]{ih95}
{Irwin}, M. \& {Hatzidimitriou}, D. 1995, \mnras, 277, 1354

\bibitem[{{Kaiser} {et~al.}(2002){Kaiser}, {Aussel}, {Burke}, {Boesgaard},
  {Chambers}, {Chun}, {Heasley}, {Hodapp}, {Hunt}, {Jedicke}, {Jewitt},
  {Kudritzki}, {Luppino}, {Maberry}, {Magnier}, {Monet}, {Onaka}, {Pickles},
  {Rhoads}, {Simon}, {Szalay}, {Szapudi}, {Tholen}, {Tonry}, {Waterson}, \&
  {Wick}}]{kais02}
{Kaiser}, N., {Aussel}, H., {Burke}, B.~E., {et~al.} 2002, in Presented at the
  Society of Photo-Optical Instrumentation Engineers (SPIE) Conference, Vol.
  4836, Society of Photo-Optical Instrumentation Engineers (SPIE) Conference
  Series, ed. {J.~A.~Tyson \& S.~Wolff}, 154--164

\bibitem[{{Keller} {et~al.}(2007){Keller}, {Schmidt}, {Bessell}, {Conroy},
  {Francis}, {Granlund}, {Kowald}, {Oates}, {Martin-Jones}, {Preston},
  {Tisserand}, {Vaccarella}, \& {Waterson}}]{kell07}
{Keller}, S.~C., {Schmidt}, B.~P., {Bessell}, M.~S., {et~al.} 2007, \pasa, 24,
  1

\bibitem[{{Kirby} {et~al.}(2010){Kirby}, {Guhathakurta}, {Simon}, {Geha},
  {Rockosi}, {Sneden}, {Cohen}, {Sohn}, {Majewski}, \& {Siegel}}]{kirb10}
{Kirby}, E.~N., {Guhathakurta}, P., {Simon}, J.~D., {et~al.} 2010, \apjs, 191,
  352

\bibitem[{{Koch} {et~al.}(2008){Koch}, {Grebel}, {Gilmore}, {Wyse}, {Kleyna},
  {Harbeck}, {Wilkinson}, \& {Wyn Evans}}]{koch08}
{Koch}, A., {Grebel}, E.~K., {Gilmore}, G.~F., {et~al.} 2008, \aj, 135, 1580

\bibitem[{{Koch} {et~al.}(2006){Koch}, {Grebel}, {Wyse}, {Kleyna}, {Wilkinson},
  {Harbeck}, {Gilmore}, \& {Evans}}]{koch06}
{Koch}, A., {Grebel}, E.~K., {Wyse}, R.~F.~G., {et~al.} 2006, \aj, 131, 895

\bibitem[{{Letarte} {et~al.}(2010){Letarte}, {Hill}, {Tolstoy}, {Jablonka},
  {Shetrone}, {Venn}, {Spite}, {Irwin}, {Battaglia}, {Helmi}, {Primas},
  {Fran{\c c}ois}, {Kaufer}, {Szeifert}, {Arimoto}, \& {Sadakane}}]{leta10}
{Letarte}, B., {Hill}, V., {Tolstoy}, E., {et~al.} 2010, \aap, 523, A17+

\bibitem[{{Martin} {et~al.}(2008){Martin}, {de Jong}, \& {Rix}}]{mart08}
{Martin}, N.~F., {de Jong}, J.~T.~A., \& {Rix}, H.-W. 2008, \apj, 684, 1075

\bibitem[{{Martin} {et~al.}(2007){Martin}, {Ibata}, {Chapman}, {Irwin}, \&
  {Lewis}}]{mart07}
{Martin}, N.~F., {Ibata}, R.~A., {Chapman}, S.~C., {Irwin}, M., \& {Lewis},
  G.~F. 2007, \mnras, 380, 281

\bibitem[{{Martinez} {et~al.}(2011){Martinez}, {Minor}, {Bullock},
  {Kaplinghat}, {Simon}, \& {Geha}}]{mart11}
{Martinez}, G.~D., {Minor}, Q.~E., {Bullock}, J., {et~al.} 2011, \apj, 738, 55

\bibitem[{{Mateo}(1998)}]{mateo98}
{Mateo}, M.~L. 1998, \araa, 36, 435

\bibitem[{{Morrison} {et~al.}(2001){Morrison}, {Olszewski}, {Mateo}, {Norris},
  {Harding}, {Dohm-Palmer}, \& {Freeman}}]{morr01}
{Morrison}, H.~L., {Olszewski}, E.~W., {Mateo}, M., {et~al.} 2001, \aj, 121,
  283

\bibitem[{{Mu{\~n}oz} {et~al.}(2006){Mu{\~n}oz}, {Majewski}, {Zaggia},
  {Kunkel}, {Frinchaboy}, {Nidever}, {Crnojevic}, {Patterson}, {Crane},
  {Johnston}, {Sohn}, {Bernstein}, \& {Shectman}}]{muno06}
{Mu{\~n}oz}, R.~R., {Majewski}, S.~R., {Zaggia}, S., {et~al.} 2006, \apj, 649,
  201

\bibitem[{{Plez}(2008)}]{plez08}
{Plez}, B. 2008, Physica Scripta Volume T, 133, 014003

\bibitem[{{Ram{\'{\i}}rez} \& {Mel{\'e}ndez}(2005)}]{rami05}
{Ram{\'{\i}}rez}, I. \& {Mel{\'e}ndez}, J. 2005, \apj, 626, 465

\bibitem[{{Rich} {et~al.}(2005){Rich}, {Corsi}, {Cacciari}, {Federici}, {Fusi
  Pecci}, {Djorgovski}, \& {Freedman}}]{rich05}
{Rich}, R.~M., {Corsi}, C.~E., {Cacciari}, C., {et~al.} 2005, \aj, 129, 2670

\bibitem[{{Robin} {et~al.}(2003){Robin}, {Reyl{\'e}}, {Derri{\`e}re}, \&
  {Picaud}}]{robin03}
{Robin}, A.~C., {Reyl{\'e}}, C., {Derri{\`e}re}, S., \& {Picaud}, S. 2003,
  \aap, 409, 523

\bibitem[{{Schiavon} {et~al.}(1997){Schiavon}, {Barbuy}, {Rossi}, \&
  {Milone}}]{schi97}
{Schiavon}, R.~P., {Barbuy}, B., {Rossi}, S.~C.~F., \& {Milone}, A. 1997, \apj,
  479, 902

\bibitem[{{Simon} \& {Geha}(2007)}]{simo07}
{Simon}, J.~D. \& {Geha}, M. 2007, \apj, 670, 313

\bibitem[{{Simon} {et~al.}(2010){Simon}, {Geha}, {Minor}, {Martinez}, {Kirby},
  {Bullock}, {Kaplinghat}, {Strigari}, {Willman}, {Choi}, {Tollerud}, \&
  {Wolf}}]{simo10}
{Simon}, J.~D., {Geha}, M., {Minor}, Q.~E., {et~al.} 2010, ArXiv e-prints

\bibitem[{{Spinrad} \& {Taylor}(1971)}]{spin71}
{Spinrad}, H. \& {Taylor}, B.~J. 1971, \apjs, 22, 445

\bibitem[{{Starkenburg} {et~al.}(2010){Starkenburg}, {Hill}, {Tolstoy},
  {Gonz{\'a}lez Hern{\'a}ndez}, {Irwin}, {Helmi}, {Battaglia}, {Jablonka},
  {Tafelmeyer}, {Shetrone}, {Venn}, \& {de Boer}}]{star10}
{Starkenburg}, E., {Hill}, V., {Tolstoy}, E., {et~al.} 2010, \aap, 513, A34+

\bibitem[{{Strigari} {et~al.}(2008){Strigari}, {Bullock}, {Kaplinghat},
  {Simon}, {Geha}, {Willman}, \& {Walker}}]{stri08}
{Strigari}, L.~E., {Bullock}, J.~S., {Kaplinghat}, M., {et~al.} 2008, \nat,
  454, 1096

\bibitem[{{Tolstoy} {et~al.}(2006){Tolstoy}, {Hill}, {Irwin}, {Helmi},
  {Battaglia}, {Letarte}, {Venn}, {Jablonka}, {Shetrone}, {Arimoto}, {Abel},
  {Primas}, {Kaufer}, {Szeifert}, {Francois}, \& {Sadakane}}]{tols06}
{Tolstoy}, E., {Hill}, V., {Irwin}, M., {et~al.} 2006, The Messenger, 123, 33

\bibitem[{{Tolstoy} {et~al.}(2009){Tolstoy}, {Hill}, \& {Tosi}}]{tht09}
{Tolstoy}, E., {Hill}, V., \& {Tosi}, M. 2009, \araa, 47, 371

\bibitem[{{Tolstoy} {et~al.}(2004){Tolstoy}, {Irwin}, {Helmi}, {Battaglia},
  {Jablonka}, {Hill}, {Venn}, {Shetrone}, {Letarte}, {Cole}, {Primas},
  {Francois}, {Arimoto}, {Sadakane}, {Kaufer}, {Szeifert}, \& {Abel}}]{tols04}
{Tolstoy}, E., {Irwin}, M.~J., {Helmi}, A., {et~al.} 2004, \apjl, 617, L119

\bibitem[{{Venn} {et~al.}(2004){Venn}, {Irwin}, {Shetrone}, {Tout}, {Hill}, \&
  {Tolstoy}}]{venn04}
{Venn}, K.~A., {Irwin}, M., {Shetrone}, M.~D., {et~al.} 2004, \aj, 128, 1177

\bibitem[{{Walker} {et~al.}(2006){Walker}, {Mateo}, {Olszewski}, {Pal}, {Sen},
  \& {Woodroofe}}]{walk06}
{Walker}, M.~G., {Mateo}, M., {Olszewski}, E.~W., {et~al.} 2006, \apjl, 642,
  L41

\bibitem[{{Walker} {et~al.}(2009{\natexlab{a}}){Walker}, {Mateo}, {Olszewski},
  {Pe{\~n}arrubia}, {Wyn Evans}, \& {Gilmore}}]{walk09b}
{Walker}, M.~G., {Mateo}, M., {Olszewski}, E.~W., {et~al.} 2009{\natexlab{a}},
  \apj, 704, 1274

\bibitem[{{Walker} {et~al.}(2009{\natexlab{b}}){Walker}, {Mateo}, {Olszewski},
  {Sen}, \& {Woodroofe}}]{walk09a}
{Walker}, M.~G., {Mateo}, M., {Olszewski}, E.~W., {Sen}, B., \& {Woodroofe}, M.
  2009{\natexlab{b}}, \aj, 137, 3109

\bibitem[{{Yi} {et~al.}(2001){Yi}, {Demarque}, {Kim}, {Lee}, {Ree}, {Lejeune},
  \& {Barnes}}]{yi01}
{Yi}, S., {Demarque}, P., {Kim}, Y.-C., {et~al.} 2001, \apjs, 136, 417

\bibitem[{{Zucker} {et~al.}(2006a){Zucker}, {Belokurov}, {Evans}, {Kleyna},
  {Irwin}, {Wilkinson}, {Fellhauer}, {Bramich}, {Gilmore}, {Newberg}, {Yanny},
  {Smith}, {Hewett}, {Bell}, {Rix}, {Gnedin}, {Vidrih}, {Wyse}, {Willman},
  {Grebel}, {Schneider}, {Beers}, {Kniazev}, {Barentine}, {Brewington},
  {Brinkmann}, {Harvanek}, {Kleinman}, {Krzesinski}, {Long}, {Nitta}, \&
  {Snedden}}]{zuck06a}
{Zucker}, D.~B., {Belokurov}, V., {Evans}, N.~W., {et~al.} 2006a, \apjl, 650,
  L41

\bibitem[{{Zucker} {et~al.}(2006b){Zucker}, {Belokurov}, {Evans}, {Wilkinson},
  {Irwin}, {Sivarani}, {Hodgkin}, {Bramich}, {Irwin}, {Gilmore}, {Willman},
  {Vidrih}, {Fellhauer}, {Hewett}, {Beers}, {Bell}, {Grebel}, {Schneider},
  {Newberg}, {Wyse}, {Rockosi}, {Yanny}, {Lupton}, {Smith}, {Barentine},
  {Brewington}, {Brinkmann}, {Harvanek}, {Kleinman}, {Krzesinski}, {Long},
  {Nitta}, \& {Snedden}}]{zuck06b}
{Zucker}, D.~B., {Belokurov}, V., {Evans}, N.~W., {et~al.} 2006b, \apjl, 643,
  L103

\end{thebibliography}

\section*{Appendix}

\begin{figure*}[]
\includegraphics[width=0.9\linewidth]{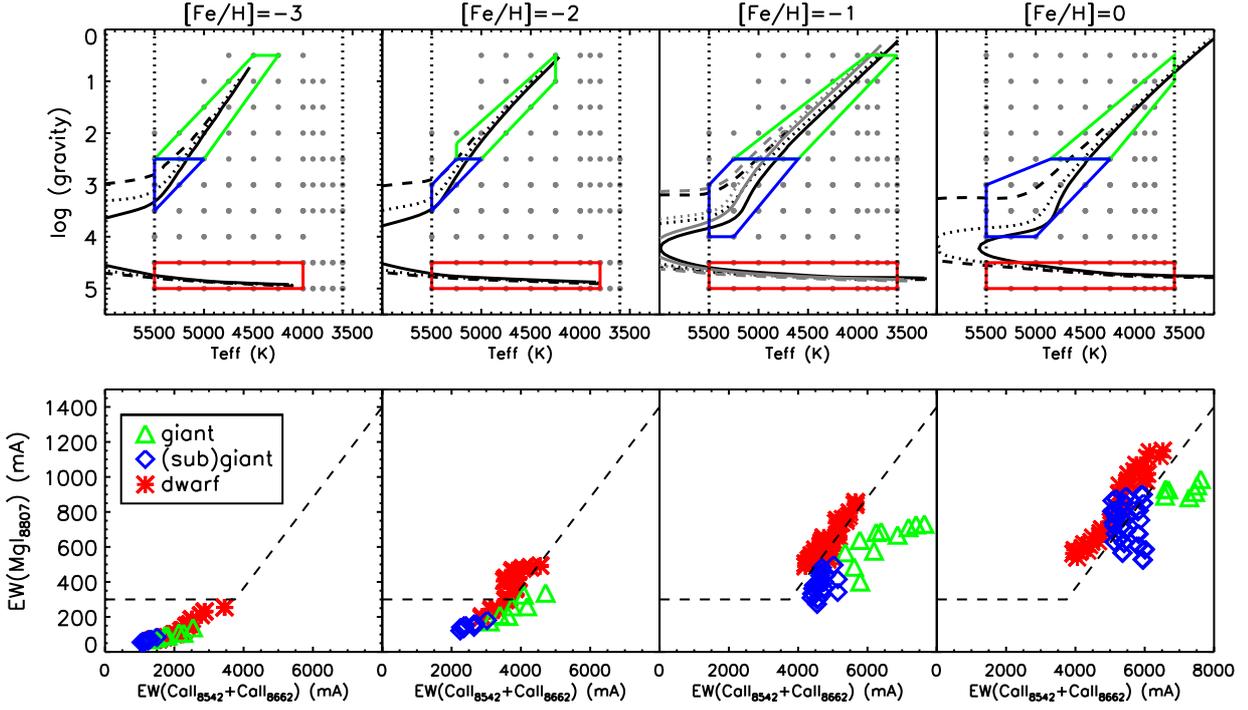}
\caption{As in Fig.~\ref{iso} but imposing an upper limit of 5500 K for the selection in temperature, rather than 5250 K.\label{iso5500}}
\end{figure*}

\begin{table*}[]
\caption{Properties of the foreground contamination towards the MW satellites Leo~IV, UMa~I and Wil~1. The columns list: the name of the galaxy (1), the number of expected MW contaminants within a solid angle of 4 deg$^2$ along the line-of-sight to the galaxy (2); the percentage of contaminants retained in the sample when using only the velocity criterion (3), the line criterion (4), when applying both (5). }
\label{tab:sel5500}
\centering
\begin{tabular}{lrrrr}
\hline
\hline 
Galaxy & N$_{cont}$ & \%(N$_{cont}$) & \%(N$_{cont}$) & \%(N$_{cont}$) \\
 & / 4 deg$^2$ & vel. crit. (3$\sigma$) & line crit. & both (3$\sigma$) \\
\hline 
Leo IV       & 8063 & 13\% & 25\% & 5\% \\
Ursa Major I & 7147 & 34\% & 17\% & 5\% \\
Willman 1    & 3806 & 36\% & 15\% & 3\% \\         
\hline
\end{tabular}
\end{table*}

Here we explore how the line 
criterion would perform down to 2 mag below the HB for Uma~I, Wil~1 and Leo~IV, i.e. those
 UFDs for which the application of the line criterion would 
considerably improve the elimination of foreground stars but 
whose faintness makes it very difficult to acquire large number of targets 
even reaching down 1 mag below the HB. 

We do so by increasing the temperature range and including models as hot as 5500~K; while this results 
into colors only milder bluer than previously considered, i.e. V-I $\sim$ 0.8, it allows to extend the magnitude range of the 
sub-giant branch boxes of about 1 mag at the fainter limit. 

Fig.~\ref{iso5500} shows that qualitatively the results of Sect.~\ref{sec:lines} continue holding: at [Fe/H]=-2 and -1 
it is possible to make a distinction between dwarfs versus giants/sub-giant stars; at [Fe/H]= 0 sub-giants overlap with both 
the locus of dwarfs and giants, but dwarfs and giants remain distinct; as before, at [Fe/H]=-3 the various types of stars 
overlap in the $\Sigma {\rm W}_{\rm CaT}$ - ${\rm EW}_{\rm Mg}$ plane. 

Obviously the actual efficiency in removing contaminants will change from what considered in Sect.~\ref{sec:others}, 
as by increasing the temperature range at the higher 
temperature end, we are not only increasing the number of contaminants but also including hotter MW stars. 
Table~\ref{tab:sel5500} shows the results when selecting stars from the Besan\c{c}on model to have 0.8 $<$V-I$<$ 2.1 and 
magnitude from the tip of the 
RGB (as done in Sect.~\ref{sec:foreg}) down to 2 mag below the HB: the fraction of contaminant stars retained in the sample by both 
velocity and line criterion increases, but the combined efficiency is still very good, with only 3-5\% of interlopers retained. 
It should be also noted that for Ursa Major~I and Willman~1 the line criterion still outperforms the velocity criterion.

\end{document}